\documentclass[aip,jap,11pt]{revtex4-1}

\usepackage{graphicx}
\usepackage{dcolumn}
\usepackage{bm}

\usepackage{braket}
\usepackage{amsmath,amssymb,amsfonts}%
\usepackage{amsthm}%

\usepackage[utf8]{inputenc}
\usepackage[T1]{fontenc}
\usepackage{mathptmx}
\usepackage{etoolbox}
\usepackage{xcolor}

\begin{document}

\title{Detection of Electron Paramagnetic Resonance of Two Electron Spins Using a Single NV Center in Diamond}

\author{Yuhang Ren}
\email{yuhangre@usc.edu}
\affiliation{Department of Physics and Astronomy, University of Southern California, Los Angeles, 90089. CA, USA}

\author{Susumu Takahashi}
\email{susumuta@usc.edu}
\affiliation{Department of Physics and Astronomy, University of Southern California, Los Angeles, 90089. CA, USA}
\affiliation{Department of Chemistry, University of Southern California, Los Angeles, 90089. CA, USA}

\date{\today}

\begin{abstract}

An interacting spin system is an excellent testbed for fundamental quantum physics and applications in quantum sensing and quantum simulation. For these investigations, detailed information of the interactions, e.g., the number of spins and their interaction strengths, is often required. In this study, we present the identification and characterization of a single nitrogen vacancy (NV) center coupled to two electron spins. In the experiment, we first identify a well-isolated single NV center and characterize its spin decoherence time. Then, we perform NV-detected electron paramagnetic resonance (EPR) spectroscopy to detect surrounding electron spins. From the analysis of the NV-EPR signal, we precisely determine the number of detected spins and their interaction strengths. Moreover, the spectral analysis indicates that the candidates of the detected spins are diamond surface spins. This study demonstrates a promising approach for the identification and characterization of an interacting spin system for realizing entangled sensing using electron spin as quantum reporters.

\end{abstract}

\keywords{EPR, Quantum Sensing, NV Center}

\maketitle

\section{Introduction}\label{sec1}

A nitrogen-vacancy (NV) center is a fluorescent impurity center existing in the diamond lattice.\cite{Science_Gruber} The unique electronic structure of the NV center allows the polarization and readout of the spin state through optical excitation and optically detected magnetic resonance (ODMR) technique.\cite{PRL_Jelezko_1,Nature_Gaebel_1,Science_CGT_1} The NV center also has a long spin coherence time at room temperature.\cite{PRL_Takahashi_1,Nature_Balasubramanian_2,science_deLange_1} It has been shown to be a promising quantum sensor, enabling measurement of an extremely small magnetic field,\cite{APL_Degen,Nature_Balasubramanian_1,Nature_Maze_1,Nature_Taylor_1} magnetic resonance spectroscopy with single spin sensitivity,\cite{Nature_Grinolds_2,Science_Shi_2,JAP_Abeywardana_1,JAP_Fortman_1,JPC_Fortman_1,APL_Fortman_1,PRB_LZP_1,PRB_Ren_1} and quantum simulations.\cite{Nature_Cai_1,ACS_Wang_1,PRB_Ju_1}

A system with interacting spins is a great platform for fundamental physics and applications in quantum information science. For example, an interacting NV spin system is a great candidate to build a quantum sensing network for quantum entanglement sensing that provides enhanced sensitivity.\cite{Science_Xie_1} Interacting spin systems also exhibit novel phases and dynamics, such as spin liquids, time crystals, and non-equilibrium dynamics.\cite{Science_Broholm,Nature_Choi_1,PRB_Francisco,JACS_Bussandri_1} NV-detected electron paramagnetic resonance (NV-EPR) spectroscopy with a single NV center is a powerful method to study nanoscale environments of interacting electron spins. NV-EPR can determine the number of the detected spins and identify the type of the detected spins. It can also determine the coupling strengths between the NV and each surrounding spin.\cite{JAP_Abeywardana_1,PRB_LZP_1} NV-EPR has been demonstrated to detect P1 centers, N2 centers and other unidentified spins.\cite{PRB_Yamamoto_1,PRB_Shi_1,PRL_Cooper_1,PRL_Rosenfeld_1,nature_Grinolds_1,PRB_LZP_1} Moreover, electron spins on diamond surfaces have been detected.\cite{PRL_Sushkov_1} A characterizable and controllable network of spins can be used for entanglement-enhanced surface quantum sensing. However, only a few demonstrations have been performed to determine the number of interacting electron spins and individual coupling strength to NV center.\cite{Science_Shi_2,PRL_Sushkov_1} Identifying and modeling such configuration remains a challenge for single NV-EPR detection.

In this work, we discuss the demonstration of NV-EPR spectroscopy of a few electron spins. We employ a single shallow NV center in diamond as a sensor and utilize a double electron-electron resonance (DEER) technique to show the detection of EPR signals of surrounding spins. The observation of the NV-EPR signal confirms the detection of spins. The type of the detected spin is analyzed from their \textit{g}-values and hyperfine splitting. Moreover, we present a simple model to determine the number of the detected spins and the strengths of their magnetic dipole interactions. We find a single NV center coupled to two electron spins with dipolar coupling $1.12\pm0.13$ MHz and $2.24\pm0.17$ MHz from the analysis of the NV-EPR signal. The presented spin system will be useful for the investigation of quantum effects that require a low number of spins\cite{NJP_XZ_1} or open quantum dynamics.\cite{SR_DSM_1} The experimental methods can be used to realize entangled quantum reporters for quantum sensing.

\section{Experimental Setup}\label{sec2}

\begin{figure}[ht]
\includegraphics[width=0.8\textwidth]{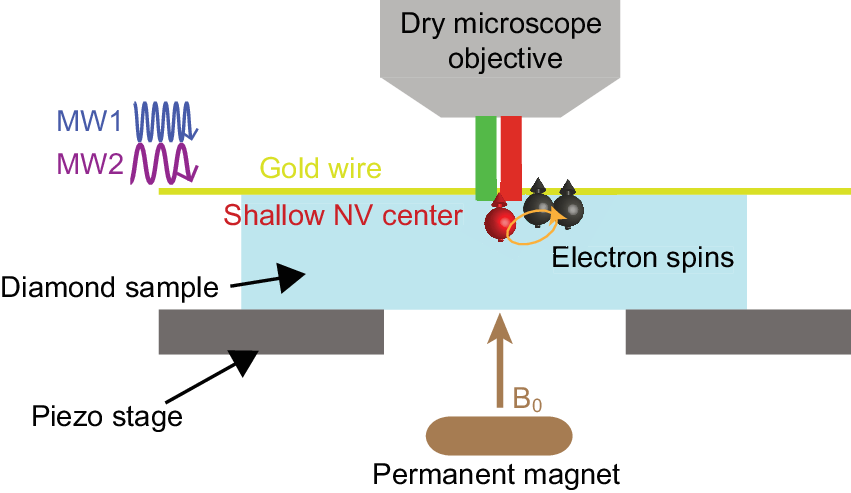}
\caption{\label{fig1}Illustration of the experimental setup. A (111)-cut diamond sample is mounted on a piezo stage. A disk magnet is used to provide static external field $B_0$, and a gold wire is attached to the sample to provide microwave excitation, MW1 and MW2. A 532 nm laser is focused onto a shallow NV near the surface of the diamond with a dry microscope objective. PL is collected through the same objective and detected with a photon detector.}
\end{figure}

In this study, we employ single NV centers in a diamond crystal sample purchased from Qnami.\cite{Qnami}  The sample is a (111)-cut, specially designed diamond. Shallow NVs ($\sim$10 nm depth) were created with Nitrogen ion implantation with the energy of 6 keV and subsequent annealing process. The sample has a unique nano-pillar structure that allows easy isolation and identification of single NV centers. We have studied four single NV centers in this sample. In the present paper, our investigation focuses on one of the single NV centers with a unique NV-EPR signal (denoted as NV1). The experimental results on the others (NV2-4) are also presented in the supplementary material. The setup of the experiment is illustrated in Fig.~\ref{fig1}. The diamond sample is mounted on an XYZ piezo stage. A permanent disk magnet is placed under the stage to provide a static magnetic field ($B_0$) along the [111] direction of the diamond lattice. For ODMR spectroscopy of the NV center, two channels of microwave (MW1 and MW2) excitation are applied using sources (Stanford SG386) and an amplifier (Mini-Circuits ZHL-15W-422-S+). Both MW1 and MW2 excitation are applied with a 20um diameter gold wire attached close to the top of the sample. A diode-pumped solid-state laser (CrystaLaser, 100 mW) and a microscope dry objective (ZEISS, 100x, NA = 0.8) are employed for the laser excitation. Photoluminescence (PL) from the NV center is collected by the same objective and redirected by a dichroic mirror towards an avalanche photon detector (Excelitas, SPCM-AQRH-13-FC). A central computer is used to control the microwave and laser pulses and analyze the photon count signal.

\section{Results and Discussion}\label{sec3}

\begin{figure}
\includegraphics[width=0.75\textwidth]{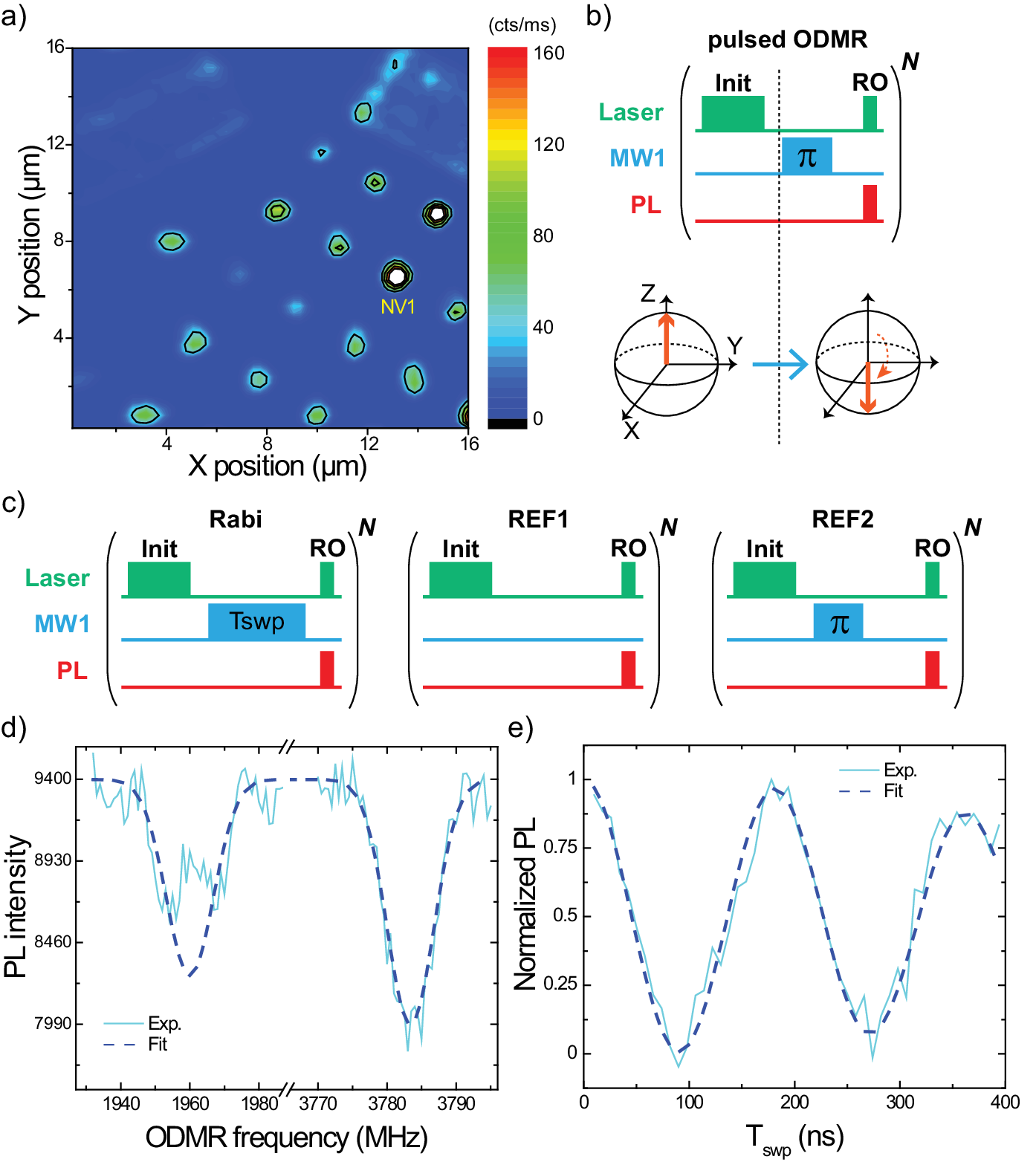}
\caption{\label{fig2}Characterization of a single NV center. (a) PL image of the diamond sample. The PL signals are from single NV centers. (b) Pulse sequence for pulsed ODMR measurement. The spin starts at $\ket{m_S = 0}$ after the initialization laser pulse (indicated as Init) and evolves under the effect of the MW1 pulse. The readout laser pulse (indicated as RO) measures the PL level according to the projected population difference. N = 100,000 is the number of pulse sequences used for averaging. (c) Pulse sequences for Rabi Oscillation measurement. Two reference channels (REF1 and REF2) are used for PL normalization. N = 100,000. (d) Pulsed ODMR data. By fitting with Gaussian peak functions, the peak frequencies are extracted to obtain the magnetic field and tilt angle from fitting with the NV Hamiltonian. (e) Rabi oscillation data. MW1 frequency is set to $\ket{m_S = 0}\leftrightarrow\ket{m_S = +1}$ transition frequency at 3783.39 MHz while the length of MW1 pulse T$_{swp}$ is swept. PL is normalized by $\frac{Rabi-REF2}{REF1-REF2}$. Rabi oscillation frequeny f of $5.50\pm0.03$ MHz and decay time $T_0$ of $0.67\pm0.08$ $\mu$s is obtained from a fit to $\frac{1}{2}\cdot (1 + exp(-(t/T_0)^2)cos(2\pi ft))$ \cite{JAP_Abeywardana_1}. The total measurement time is around 5 minutes. }
\end{figure}

The NV-EPR experiment starts with the characterization of the external magnetic field and the coherent manipulation capability of the NV centers. After identifying single NV centers using the auto-correlation experiment and ODMR spectroscopy,\cite{JAP_Abeywardana_1} we focus on our investigation into each single NV and perform a pulsed ODMR measurement to determine the magnetic field strength and tilt angle. Fig.~\ref{fig2}(a) shows a PL image of the sample and the studied NV1 is indicated in the image. Next, we perform pulsed ODMR measurements. As shown in Fig.~\ref{fig2}(b), the laser pulse with a duration of 5 $\mu$s initialize (Init) the NV state to $\ket{m_S = 0}$. After a short delay ($\sim$1 $\mu$s), a MW1 $(\pi)_Y$ pulse is applied to drive a rotation of the spin in the Bloch's sphere along the x-axis (see Fig.~\ref{fig2}(b)), from brighter $\ket{m_S = 0}$ to darker $\ket{m_S = \pm 1}$ spin states. Afterward a short readout (RO) laser pulse with a duration of 300 ns project NV's coherent spin state into the population difference, which is measured by the PL intensity, resulting in observation of ODMR signals at the $\ket{m_S = 0}$ $\leftrightarrow$ $\ket{m_S = +1}$ and $\ket{m_S = 0}$ $\leftrightarrow$ $\ket{m_S = -1}$ transition frequencies. Fig.~\ref{fig2}(d) shows the PL intensity as a function of the MW1 frequency. Pulse sequence is averaged over 100,000 times in around 7 minutes to obtain the spectrum of each peak. By fitting the two frequencies, $1960.00 \pm 6.78$ MHz and $3783.39 \pm 3.39$ MHz, to the following NV Hamiltonian, we obtain the magnetic field strength ($B_0$) and the tilt angle ($\theta$) of the magnetic field in respect to the NV axis,

\begin{equation}
\label{eq1}
    H = \gamma_{NV} B_0 \left( sin\theta \cdot S_x + cos\theta \cdot S_z \right) + DS_z^2,
\end{equation}
where $\gamma_{NV}$ is the gyromagnetic ratio of the NV spin (28.024 GHz/Tesla). $S = (S_x,S_y,S_z)$ are the spin operators of $S=1$. $D$ is the NV zero-field energy splitting (2.87 GHz). As can be seen in Fig.~\ref{fig2}(d), the experimental data and fit result are in fair agreement. From the fit result, we obtained the magnetic field strength of $B_0 = 32.59 \pm 0.02$ mT and the tilt angle of $\theta = 3.5 \pm 0.8$ degrees. Next, we perform a measurement of Rabi oscillations. The pulse sequences of the experiment are shown in Fig.~\ref{fig2}(c). After the laser pulse initialization, an MW1 $(\pi)_Y$ pulse at the NV1 Larmor frequency of 3783.39 MHz (corresponding to the transition $\ket{m_S = 0}$ $\leftrightarrow$ $\ket{m_S = +1}$) with a duration of $T_{swp}$ is applied. Then, the resultant population change is readout with the RO laser pulse. Two additional reference channels (REF1 and REF2) are used for data normalization. REF1 corresponds to the maximum PL detected in Rabi measurement, and REF2 corresponds to the minimum PL. The normalized PL is $(Rabi-REF2)/(REF1-REF2)$. This normalization procedure will reduce degrees of freedom in data fitting. As shown in Fig.~\ref{fig2}(e), the normalized PL shows oscillations as the MW1 pulse length is varied. These are the Rabi oscillations that prove the coherent manipulation of the NV spin states. From the period of the oscillations, the duration of the $\pi$, $\pi/2$, and $3\pi/2$ pulses can be determined. In the present case, we obtained the $\pi$, $\pi/2$, and $3\pi/2$ pulse duration to be 46, 92, and 138 ns, respectively.

\begin{figure}[ht]
\includegraphics[width=0.9\textwidth]{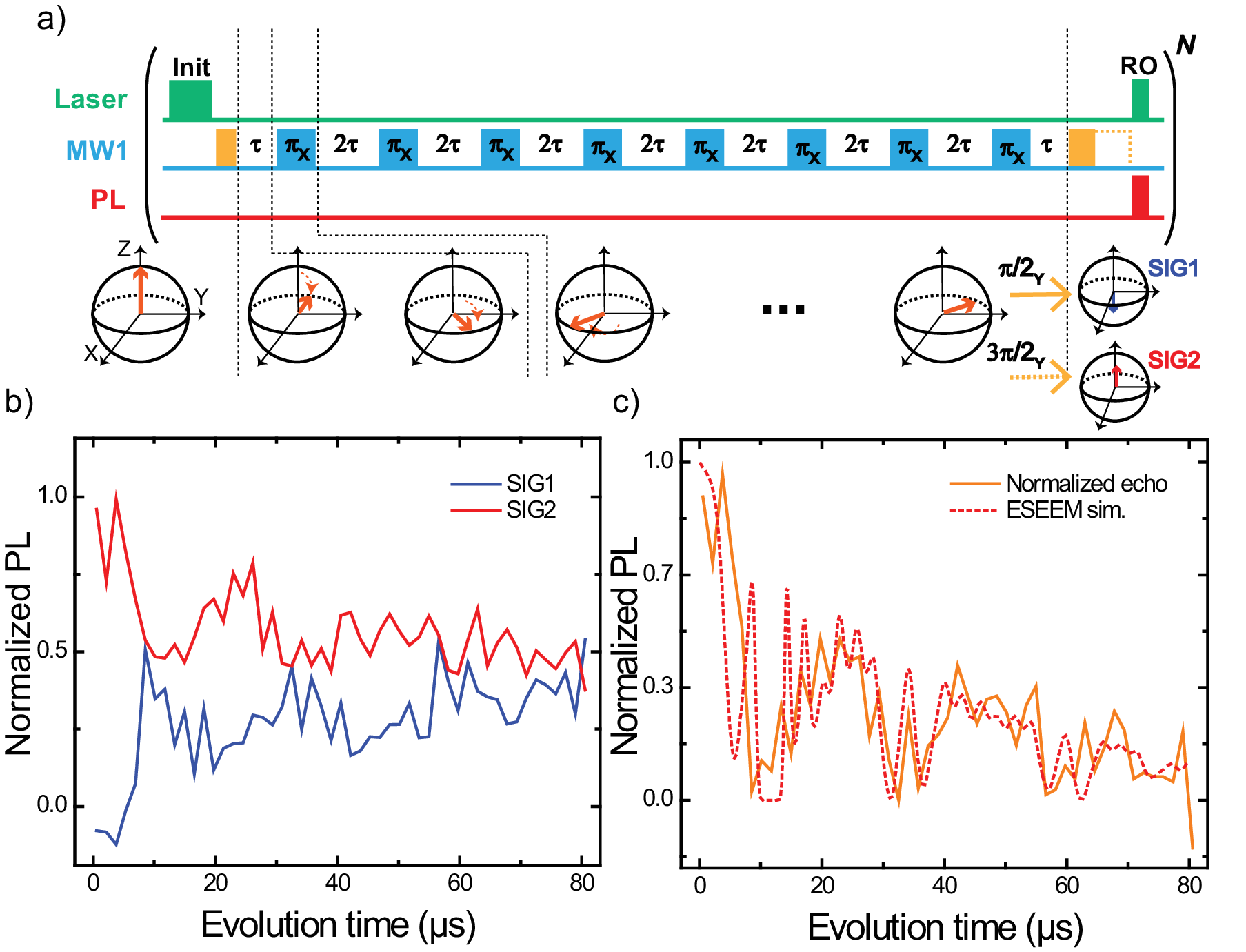}
\caption{\label{fig3}CPMG-8 sequence and data. (a) CPMG-8 sequence. The yellow and blue MW1 pulses are with X/Y phases. SIG1 channel uses a $(\pi/2)_Y$ pulse to project the phase into population difference, while SIG2 channel uses a $(3\pi/2)_Y$ pulse (indicated with a dashed yellow box). Laser Init and RO pulse lengths are 5 $\mu$s and 300 ns. MW1 $\pi/2$, $\pi$ and $3\pi/2$ pulse lengths are 46, 92 and 138 ns. N = 220,000. (b) CPMG-8 data showing normalized PL. The evolution time is calculated to include all evolution periods, namely 16$\tau$. The measurement time for this spectrum is around 32 minutes. (c) The normalized signal (SIG2 - SIG1) is fit to a simulation including $^{13}$C bath and a nearby $^{13}$C spin ESEEM effect with $T_2$ decay. We extract a $T_2$ time of $38 \pm 3$ $\mu$s.}
\end{figure}

Next, we characterize the spin decoherence time ($T_2$) using a Carr-Purcell-Meiboom-Gill (CPMG) sequence. The particular sequence used here consists of eight $\pi$ pulses, which is referred to later as a CPMG-8 sequence (see Fig.~\ref{fig3}(a)). In the CPMG-8 experiment, after the preparation of the superposition state $\frac{1}{\sqrt{2}}(\ket{+1}+\ket{0})$ by the $(\pi/2)_Y$ pulse, a chain of $(\pi)_X$ pulses are applied at every 2$\tau$ to keep the coherence state for an extended period. The resultant coherent state is mapped into the $\ket{m_s=0}$ state with the application of the second $(\pi/2)_Y$ pulse (SIG1). In addition, in a different signal channel, we use a $(3\pi/2)_Y$ pulse instead of the $(\pi/2)_Y$ pulse to map the state into the $\ket{m_s=1}$ state (SIG2). Two signal channels are used to improve the signal-to-noise ratio (SNR) by removing systematic noise and doubling signal contrast. PL intensity is normalized using the same method used in the Rabi oscillation experiment. As seen in Fig.~\ref{fig3}(b), an oscillating and decaying difference in the normalized PL is observed. These features in the echo signal are attributed to the electron spin echo envelope modulation (ESEEM) effect due to hyperfine interaction with nearby nuclear spins.\cite{Childress_2006} As shown in Fig.~\ref{fig3}(c), we found that the simulation of the ESEEM signal can explain the observed signal taken with the CPMG sequence. In the simulation, we consider bath $^{13}$C nuclear spins and a $^{13}$C individual spin with the hyperfine couplings of $A=0.314$ MHz and $B = 2.827$ MHz. We also obtained the spin decoherence time ($T_2$) to be $38 \pm 3$ $\mu$s. This detection of the individual $^{13}$C nuclear spin was observed only with NV1, not with other NVs studied in this work. The details of the simulation are included in the supplementary material.

\begin{figure}[ht]
\includegraphics[width=0.9\textwidth]{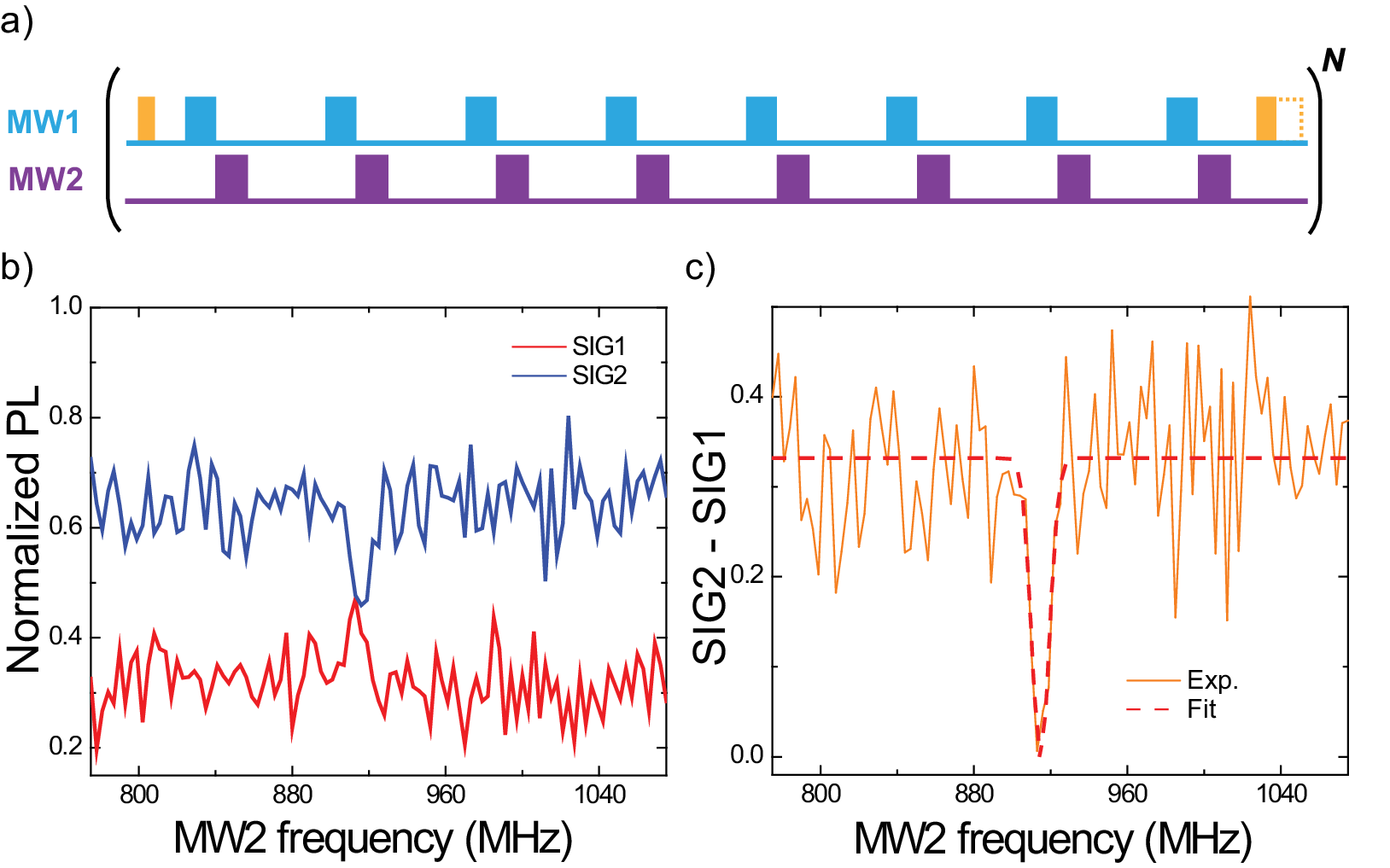}
\caption{\label{fig4}NV-EPR sequence and data. (a) The highlight of the CPMG-DEER sequence showing only the MW1 and MW2 pulses, indicated by blue and purple boxes, respectively. MW1 $\pi/2$ and $\pi$ pulse lengths are 46 and 92 ns, and MW2 $\pi$ pulse length is 92 ns. $\tau$ time is set to 1.28 $\mu$s and the total sensing time is 20.864 $\mu$s. N = 1,325,000. (b) CPMG-DEER data. Both SIG1 and SIG2 channels are used. The total measurement time is around 320 minutes. The signals are normalized in the same way as the CPMG experiment. (c) By fitting the difference (SIG2-SIG1) to a Gaussian peak function, we obtain the narrow DEER spectrum centered at $914.7 \pm 0.9$ MHz, with no sign of hyperfine coupling in the vicinity.}
\end{figure}

Finally, we perform NV-detected EPR (NV-EPR) spectroscopy to detect electron spins surrounding NV1. In the NV-EPR experiment, we use a DEER sequence shown in Fig.~\ref{fig4}(a). We use two microwaves (MW1 and MW2) to coherently control the NV spin and target environment spins, respectively. The first microwave (MW1) is fixed at NV Lamour frequency. The second microwave (MW2) is swept in frequency, and when the frequency of MW2 is at the Lamour frequency of the target spins, the MW2 pulse flips the target spins. As can be seen from Fig.~\ref{fig4}(a), the sequence consists of the CPMG component for the NV spin and the $\pi$ pulses for the target spins (denoted as CPMG-DEER). In the CPMG-DEER experiment, first, the NV's spin state is prepared into $\ket{\psi_0} = \frac{1}{\sqrt{2}}(\ket{+1}+\ket{0})$ state. The spin state is sensitive to small changes in the external magnetic field $B_0$, which in our case is the total dipole field $B_{dip}$ from surrounding electron spins. In the present case, with the application of the $\pi$ pulses with MW2 (See Fig.~\ref{fig4}(a)), phase shifts due to changes of the magnetic field are accumulated to the NV coherent state, i.e., $\ket{\psi_0} \rightarrow \ket{\psi} = \frac{1}{\sqrt2} (\ket{+1} + e^{i\delta\phi}\ket{0})$. The phase shifts can be modeled as,

\begin{equation}
\label{eq2}
    \delta \phi = \frac{\gamma_{NV}}{\hbar} \left[ \int_0^{\tau} B_{dip}(t) dt - \int_{\tau}^{3\tau} B_{dip}(t) dt + ... + \int_{13\tau}^{15\tau} B_{dip}(t) dt - \int_{15\tau}^{16\tau} B_{dip}(t) dt \right],
\end{equation}
where $\hbar$ is the reduced Planck constant. The pulse length is relatively short compared to $\tau$ time, so the evolution during pulses is neglected in the equation. The spin flip-flop not due to pulse excitation is also neglected.\cite{PRB_Stepanov_1} $B_{dip}(t) = \frac{\mu_0}{4\pi}\sum_i g_e\mu_B(3cos^2\theta_i-1)\sigma_i/r^3_i$ is the total magnetic field generated by the target spin(s) at the NV. $g_e$ is the $g$-value of the target spin. $\theta_i$ and $r_i$ are the angle and distance between the NV spin and target spin. $B_{dip}$ is assumed to be constant throughout one evolution period. When the MW2 pulse is not on resonance, it does not drive any state evolution. As a result, the odd and even terms in Eqn.~\ref{eq2} are canceled. On the other hand, when the frequency of MW2 is on resonance, the MW2 pulse flips the target spins. This spin flip changes the sign of $B_{dip}$ in even terms of Eqn.\ref{eq2}, i.e., $B_{dip} \rightarrow -B_{dip}$, allowing the phase to accumulate over the total sensing period. Namely, Eqn.\ref{eq2} can be rewritten as,

\begin{equation}
\label{eq3}
\begin{split}
    \text{On Resonance: } \delta \phi &= \frac{g_{NV}\mu_B}{\hbar} \left[ \int_0^{\tau} B_{dip} dt + \int_{\tau}^{3\tau} B_{dip} dt + ... \right] \\
    &= \frac{g_{NV}\mu_B}{\hbar} \int_0^{16\tau} B_{dip} dt, \\
    \text{Off Resonance: } \delta \phi &= 0.
\end{split}
\end{equation}
It is worth noting that even though the bath spins are not polarized to be in the same state in each measurement, we can still observe the NV-EPR signal since it is due to the statistical average of the spin polarization instead of a thermal polarization. In general, the NV-EPR signal due to a phase shift of $\delta\phi$ is given by,

\begin{equation}
\label{eq4}
    I_{\text{NVEPR}} = \frac{1}{2} \left( 1 + \langle cos(\delta \phi) \rangle \right),
\end{equation}
where $\left<...\right>$ represents the statistical average. In the present case, we consider a case where the phase shifts are caused by the flips of target spins with the application of the MW2 pulses ($T_{MW2}$). Then, the NV-EPR signal is given by,\cite{PRB_Stepanov_1}

\begin{equation}
\begin{split}
\label{eq5}
    I_{\text{NVEPR}} &= \frac{1}{2} + \frac{1}{2} exp(-(\frac{t}{T_0})^2) \left\langle cos\left( \frac{\mu_0}{4\pi} \frac{\mu_B^2g_{NV}g_e T_{MW2}}{\hbar} \sum_j \frac{(3cos^2\theta_j-1)\sigma_j}{r_j^3} \right) \right\rangle_{\sigma_j} \\
    &= \frac{1}{2} + \frac{1}{2} exp(-(\frac{t}{T_0})^2) \left\langle cos\left( \sum_j \omega_j T_{MW2} \right) \right\rangle_{\sigma_j},
\end{split}
\end{equation}
where $\sigma_j$ is +1 or -1 representing $\ket{\uparrow}$ or $\ket{\downarrow}$, respectively. $\left<...\right>_{\sigma_j}$ here denotes the statistical average of all conditions with different combinations of $\sigma_j$ values. $\mu_0$ is the vacuum magnetic permeability. $T_{MW2}$ is the MW2 pulse duration. $\omega_j$ is the Rabi frequency associated with dipole interaction strength. Similar to the experiment in Fig.~\ref{fig2}(d), the Gaussian decay is included to take into account a decay in Rabi oscillations\cite{JAP_Abeywardana_1} where $T_0$ is the decay constant. When the NV-EPR signal is from a single spin, the oscillations are given by a single sinusoidal function. When it is from two spins, the oscillations are given by a sum of two sinusoidal functions, etc. Therefore, the analysis of the oscillations can be used to determine the number of spins coupled to the NV center as well as their coupling strength.

\begin{figure}[ht]
\includegraphics[width=0.9\textwidth]{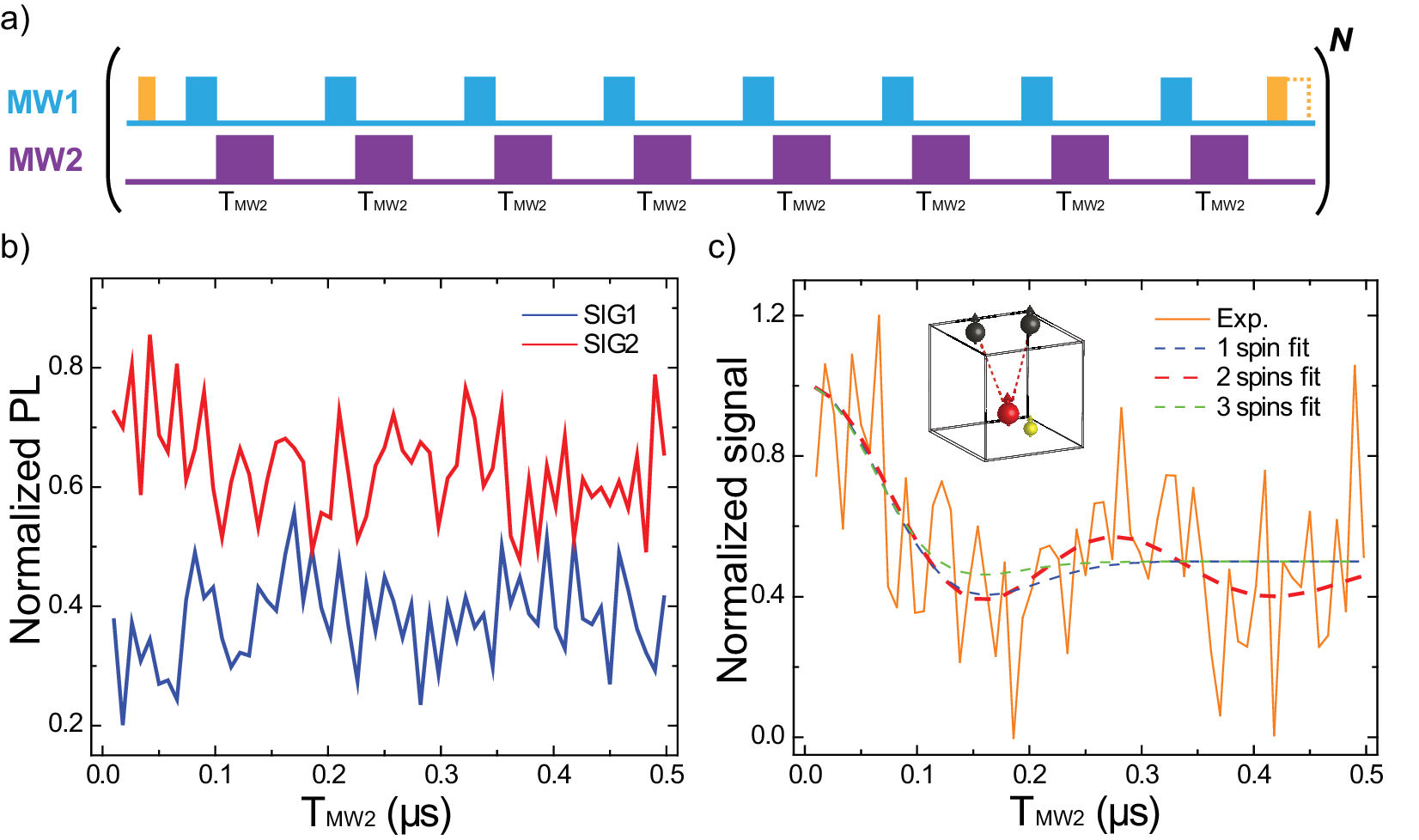}
\caption{\label{fig5}Rabi oscillation of NV-EPR signal. (a) DEER-Rabi sequence. MW1 $\pi/2$ and $\pi$ pulse lengths are 46 and 92 ns. The length of MW2 pulses is swept. N = 1,260,000. (b) Rabi oscillation of the NV-EPR signal. 
The measurement time is around 180 minutes. Both SIG1 and SIG2 PL intensities are normalized the same way as in previous experiments. (c) The difference (SIG2-SIG1) is normalized from 0 to 1 and then fit to a simple NV-EPR model, where the two electron spins configuration gives the best fit. We obtained the interaction strength between NV1 and the electron spins to be $\omega_1 = 2\pi \times (1.12 \pm 0.13)$ MHz and $\omega_2 = 2\pi \times (2.24 \pm 0.17)$ MHz, respectively.}
\end{figure}

Next, we introduce another NV-EPR experiment to determine the number of spins and strength of interactions detected in the NV-EPR signal. The sequence is shown in Fig.~\ref{fig5}(a). This sequence is modified from the CPMG-DEER sequence to measure the Rabi oscillations of the NV-EPR signal, so we call it a DEER-Rabi sequence. In the DEER-Rabi sequence, the MW2 frequency is fixed at the EPR signal, and the pulse length ($T_{MW2}$) is varied while keeping the total echo evolution time fixed. Both SIG1 and SIG2 channels are used as well for better SNR. Data is shown in Fig.~\ref{fig5}(b). The normalized DEER-Rabi signal (SIG2-SIG1) shows a strong oscillation in the population recovery (shown in Fig.~\ref{fig5}(c)). The observed oscillations are due to the change of dipole field from spins felt by the NV coherence state. The MW2 pulses rotate the spins according to the pulse length, changing the strength of the field along the NV axis. The accumulated phase shifts, therefore, undergo oscillations. To analyze the signal (shown in Fig.~\ref{fig5}(c)) with Eqn.~\ref{eq5}, we consider three situations where the number of target spins ranges from 1 to 3. In the fit, the dipole interaction strength and decay constant are treated as fitting parameters. We set the range for interaction strength to be 2-312.5 MHz. The lower bound is taken to have one full oscillation in the total recorded pulse length, and the higher bound is taken to have one full oscillation for every four data points. This follows from the frequency resolution being 1/(measurement time) and maximum frequency being (sample rate)/2 for a discrete Fourier transform analysis. The fitted lines are plotted as dotted lines in Fig.~\ref{fig5}(c). The adjusted R-square value ($1-(1-\frac{\sum (y_i-\hat{y}_i)^2}{\sum (y_i-\bar{y})^2} \frac{n-1}{n-k-1})$) is used, where $k$ = 3. In general the value is between 0 and 1, and the closer case to 1 means better agreement. The two electron spin model has the best fit with the adjusted R-square value of 0.314, while one spin model gives 0.269 and three spin model gives 0.233. From the fitting result, two interaction strengths are $\omega_1 = 2\pi \times (1.12 \pm 0.13)$ MHz and $\omega_2 = 2\pi \times (2.24 \pm 0.17)$ MHz. The time constant $T_0$ is fit to be $0.34\pm0.06$ $\mu$s. The inset of Fig.~\ref{fig5}(c) summarizes the detected two electron spins and one nuclear spin from the present study with NV1.

Finally, we discuss the origin of the NV-EPR signal observed in the present investigation (see Fig.~\ref{fig4}(c) and (d)). The signal is a single peak of coherence population difference centered at $914.7 \pm 0.9$ MHz with a width of $9 \pm 2$ MHz. No signature of the hyperfine splitting is observed. We attribute this signal to an electron spin with $S = 1/2$ and $g = 2.009\pm0.003$. Since there are no signatures of hyperfine peaks, the signal is not from P1 centers, which are commonly observed in diamonds with a high nitrogen concentration. The identification of the detected spins remains unclear, but based on the $S$ and $g$-value and lack of the hyperfine splitting, possible candidates of the observed spins are the tri-Nitrogen  W21 center\cite{RPP_Loubser_1,Baker} and surface spins.\cite{nature_Grinolds_1,JCP_Peng_1,JAP_Peng_1}

\section{Summary}\label{sec4}

In summary, we demonstrated the detection of EPR from two electron spins using a single NV center in diamond. CPMG-DEER and DEER-Rabi sequences are used to study the spin interactions. The observed NV-EPR signal is in the absence of strong hyperfine peaks, and further measurement of Rabi oscillations of the signal suggests the interaction is of a single NV center with two electron spins. The physical system and method presented in this paper allow easy identification of a single defect center coupled to a small number of spins and characterization of the interaction strength. The analysis of the NV-EPR signal presented can be a method to distinguish the number of interacting spins up to five, which is limited by the detection time window of the oscillations. 

The non-equal but measurable interaction strength is favorable in studying non-equilibrium quantum dynamics. Investigation of quantum phase dynamics, like spin liquids or discrete time crystalline (DTC), requires interacting spin systems with disordered structures.\cite{Science_Broholm,JACS_Bussandri_1} Cu$^{2+}$ ions formed kagome lattice\cite{RMP_Norman} and Ru$^{3+}$ ions formed honeycomb lattice\cite{Nature_Takagi_1} has been actively researched materials for realization of spin liquid. This multi electron spin system could be a material to look for these emergent quantum effects with the similarity of being a spin 1/2 magnetic disordered type of material. A chain of 9 $^{13}$C nuclear spins in diamond has been shown to be a programmable spin-based DTC.\cite{Science_Randall_1} The characterizable electron spin system in this study is potentially a smaller-sized system to realize DTC. A good knowledge of the interaction strengths in a spin system will also be significant in better characterization of non-classical correlations, namely quantum entanglement, which can be key in modeling open quantum dynamics and developing entangled quantum sensing. Quantum-enhanced NMR with a 9.4-fold sensitivity increase has been demonstrated by entangling 9 proton spins with a $^{31}$P spin.\cite{Science_Jonathan} The electron spins detected in this study can also be simultaneously controlled and potentially achieve entanglement enhancement beyond the standard quantum limit. In addition, further study can be done to reveal more information about the nature of the spin species and improve the understanding of the decoherence of shallow NV centers, paving the pathway to higher standard engineering for quantum sensing applications. 

\section{Supplementary Material}

The supplementary material includes a discussion and simulation of the ESEEM effect observed in the experiment and additional measurement results on the other NV centers.

\begin{acknowledgments}

This work was supported by the National Science Foundation (ECCS-2204667, CHE-2404463 and CHE-2004252 with partial co-funding from the Quantum Information Science program in the Division of Physics), the USC Dornsife, the USC Anton B. Burg Foundation, and the Searle scholars program (ST).

\end{acknowledgments}

\section*{Data Availability}

The data that support the findings of this study are available from the corresponding author upon reasonable request.

\bibliography{bibliography}

\begin{thebibliography}{46}%
\makeatletter
\providecommand \@ifxundefined [1]{%
 \@ifx{#1\undefined}
}%
\providecommand \@ifnum [1]{%
 \ifnum #1\expandafter \@firstoftwo
 \else \expandafter \@secondoftwo
 \fi
}%
\providecommand \@ifx [1]{%
 \ifx #1\expandafter \@firstoftwo
 \else \expandafter \@secondoftwo
 \fi
}%
\providecommand \natexlab [1]{#1}%
\providecommand \enquote  [1]{``#1''}%
\providecommand \bibnamefont  [1]{#1}%
\providecommand \bibfnamefont [1]{#1}%
\providecommand \citenamefont [1]{#1}%
\providecommand \href@noop [0]{\@secondoftwo}%
\providecommand \href [0]{\begingroup \@sanitize@url \@href}%
\providecommand \@href[1]{\@@startlink{#1}\@@href}%
\providecommand \@@href[1]{\endgroup#1\@@endlink}%
\providecommand \@sanitize@url [0]{\catcode `\\12\catcode `\$12\catcode `\&12\catcode `\#12\catcode `\^12\catcode `\_12\catcode `\%12\relax}%
\providecommand \@@startlink[1]{}%
\providecommand \@@endlink[0]{}%
\providecommand \url  [0]{\begingroup\@sanitize@url \@url }%
\providecommand \@url [1]{\endgroup\@href {#1}{\urlprefix }}%
\providecommand \urlprefix  [0]{URL }%
\providecommand \Eprint [0]{\href }%
\providecommand \doibase [0]{http://dx.doi.org/}%
\providecommand \selectlanguage [0]{\@gobble}%
\providecommand \bibinfo  [0]{\@secondoftwo}%
\providecommand \bibfield  [0]{\@secondoftwo}%
\providecommand \translation [1]{[#1]}%
\providecommand \BibitemOpen [0]{}%
\providecommand \bibitemStop [0]{}%
\providecommand \bibitemNoStop [0]{.\EOS\space}%
\providecommand \EOS [0]{\spacefactor3000\relax}%
\providecommand \BibitemShut  [1]{\csname bibitem#1\endcsname}%
\let\auto@bib@innerbib\@empty
\bibitem [{\citenamefont {Gruber}\ \emph {et~al.}(1997)\citenamefont {Gruber}, \citenamefont {Dräbenstedt}, \citenamefont {Tietz}, \citenamefont {Fleury}, \citenamefont {Wrachtrup},\ and\ \citenamefont {von Borczyskowski}}]{Science_Gruber}%
  \BibitemOpen
  \bibfield  {author} {\bibinfo {author} {\bibfnamefont {A.}~\bibnamefont {Gruber}}, \bibinfo {author} {\bibfnamefont {A.}~\bibnamefont {Dräbenstedt}}, \bibinfo {author} {\bibfnamefont {C.}~\bibnamefont {Tietz}}, \bibinfo {author} {\bibfnamefont {L.}~\bibnamefont {Fleury}}, \bibinfo {author} {\bibfnamefont {J.}~\bibnamefont {Wrachtrup}}, \ and\ \bibinfo {author} {\bibfnamefont {C.}~\bibnamefont {von Borczyskowski}},\ }\href {\doibase 10.1126/science.276.5321.2012} {\bibfield  {journal} {\bibinfo  {journal} {Science}\ }\textbf {\bibinfo {volume} {276}},\ \bibinfo {pages} {2012} (\bibinfo {year} {1997})}\BibitemShut {NoStop}%
\bibitem [{\citenamefont {Jelezko}\ \emph {et~al.}(2004)\citenamefont {Jelezko}, \citenamefont {Gaebel}, \citenamefont {Popa}, \citenamefont {Gruber},\ and\ \citenamefont {Wrachtrup}}]{PRL_Jelezko_1}%
  \BibitemOpen
  \bibfield  {author} {\bibinfo {author} {\bibfnamefont {F.}~\bibnamefont {Jelezko}}, \bibinfo {author} {\bibfnamefont {T.}~\bibnamefont {Gaebel}}, \bibinfo {author} {\bibfnamefont {I.}~\bibnamefont {Popa}}, \bibinfo {author} {\bibfnamefont {A.}~\bibnamefont {Gruber}}, \ and\ \bibinfo {author} {\bibfnamefont {J.}~\bibnamefont {Wrachtrup}},\ }\href {\doibase 10.1103/PhysRevLett.92.076401} {\bibfield  {journal} {\bibinfo  {journal} {Phys. Rev. Lett.}\ }\textbf {\bibinfo {volume} {92}},\ \bibinfo {pages} {076401} (\bibinfo {year} {2004})}\BibitemShut {NoStop}%
\bibitem [{\citenamefont {Gaebel}\ \emph {et~al.}(2006)\citenamefont {Gaebel}, \citenamefont {Domhan}, \citenamefont {Popa}, \citenamefont {Wittmann}, \citenamefont {Neumann}, \citenamefont {Jelezko}, \citenamefont {Rabeau}, \citenamefont {Stavrias}, \citenamefont {Greentree}, \citenamefont {Prawer}, \citenamefont {Meijer}, \citenamefont {Twamley}, \citenamefont {Hemmer},\ and\ \citenamefont {Wrachtrup}}]{Nature_Gaebel_1}%
  \BibitemOpen
  \bibfield  {author} {\bibinfo {author} {\bibfnamefont {T.}~\bibnamefont {Gaebel}}, \bibinfo {author} {\bibfnamefont {M.}~\bibnamefont {Domhan}}, \bibinfo {author} {\bibfnamefont {I.}~\bibnamefont {Popa}}, \bibinfo {author} {\bibfnamefont {C.}~\bibnamefont {Wittmann}}, \bibinfo {author} {\bibfnamefont {P.}~\bibnamefont {Neumann}}, \bibinfo {author} {\bibfnamefont {F.}~\bibnamefont {Jelezko}}, \bibinfo {author} {\bibfnamefont {J.~R.}\ \bibnamefont {Rabeau}}, \bibinfo {author} {\bibfnamefont {N.}~\bibnamefont {Stavrias}}, \bibinfo {author} {\bibfnamefont {A.~D.}\ \bibnamefont {Greentree}}, \bibinfo {author} {\bibfnamefont {S.}~\bibnamefont {Prawer}}, \bibinfo {author} {\bibfnamefont {J.}~\bibnamefont {Meijer}}, \bibinfo {author} {\bibfnamefont {J.}~\bibnamefont {Twamley}}, \bibinfo {author} {\bibfnamefont {P.~R.}\ \bibnamefont {Hemmer}}, \ and\ \bibinfo {author} {\bibfnamefont {J.}~\bibnamefont {Wrachtrup}},\ }\href {\doibase 10.1038/nphys318} {\bibfield  {journal} {\bibinfo  {journal} {Nat. Phys.}\ }\textbf
  {\bibinfo {volume} {2}},\ \bibinfo {pages} {408} (\bibinfo {year} {2006})}\BibitemShut {NoStop}%
\bibitem [{\citenamefont {Childress}\ \emph {et~al.}(2006{\natexlab{a}})\citenamefont {Childress}, \citenamefont {Dutt}, \citenamefont {Taylor}, \citenamefont {Zibrov}, \citenamefont {Jelezko}, \citenamefont {Wrachtrup}, \citenamefont {Hemmer},\ and\ \citenamefont {Lukin}}]{Science_CGT_1}%
  \BibitemOpen
  \bibfield  {author} {\bibinfo {author} {\bibfnamefont {L.}~\bibnamefont {Childress}}, \bibinfo {author} {\bibfnamefont {M.~V.~G.}\ \bibnamefont {Dutt}}, \bibinfo {author} {\bibfnamefont {J.~M.}\ \bibnamefont {Taylor}}, \bibinfo {author} {\bibfnamefont {A.~S.}\ \bibnamefont {Zibrov}}, \bibinfo {author} {\bibfnamefont {F.}~\bibnamefont {Jelezko}}, \bibinfo {author} {\bibfnamefont {J.}~\bibnamefont {Wrachtrup}}, \bibinfo {author} {\bibfnamefont {P.~R.}\ \bibnamefont {Hemmer}}, \ and\ \bibinfo {author} {\bibfnamefont {M.~D.}\ \bibnamefont {Lukin}},\ }\href {\doibase 10.1126/science.1131871} {\bibfield  {journal} {\bibinfo  {journal} {Science}\ }\textbf {\bibinfo {volume} {314}},\ \bibinfo {pages} {281} (\bibinfo {year} {2006}{\natexlab{a}})}\BibitemShut {NoStop}%
\bibitem [{\citenamefont {Takahashi}\ \emph {et~al.}(2008)\citenamefont {Takahashi}, \citenamefont {Hanson}, \citenamefont {van Tol}, \citenamefont {Sherwin},\ and\ \citenamefont {Awschalom}}]{PRL_Takahashi_1}%
  \BibitemOpen
  \bibfield  {author} {\bibinfo {author} {\bibfnamefont {S.}~\bibnamefont {Takahashi}}, \bibinfo {author} {\bibfnamefont {R.}~\bibnamefont {Hanson}}, \bibinfo {author} {\bibfnamefont {J.}~\bibnamefont {van Tol}}, \bibinfo {author} {\bibfnamefont {M.~S.}\ \bibnamefont {Sherwin}}, \ and\ \bibinfo {author} {\bibfnamefont {D.~D.}\ \bibnamefont {Awschalom}},\ }\href {\doibase 10.1103/PhysRevLett.101.047601} {\bibfield  {journal} {\bibinfo  {journal} {Phys. Rev. Lett.}\ }\textbf {\bibinfo {volume} {101}},\ \bibinfo {pages} {047601} (\bibinfo {year} {2008})}\BibitemShut {NoStop}%
\bibitem [{\citenamefont {Balasubramanian}\ \emph {et~al.}(2009)\citenamefont {Balasubramanian}, \citenamefont {Neumann}, \citenamefont {Twitchen}, \citenamefont {Markham}, \citenamefont {Kolesov}, \citenamefont {Mizuochi}, \citenamefont {Isoya}, \citenamefont {Achard}, \citenamefont {Beck}, \citenamefont {Tissler}, \citenamefont {Jacques}, \citenamefont {Hemmer}, \citenamefont {Jelezko},\ and\ \citenamefont {Wrachtrup}}]{Nature_Balasubramanian_2}%
  \BibitemOpen
  \bibfield  {author} {\bibinfo {author} {\bibfnamefont {G.}~\bibnamefont {Balasubramanian}}, \bibinfo {author} {\bibfnamefont {P.}~\bibnamefont {Neumann}}, \bibinfo {author} {\bibfnamefont {D.}~\bibnamefont {Twitchen}}, \bibinfo {author} {\bibfnamefont {M.}~\bibnamefont {Markham}}, \bibinfo {author} {\bibfnamefont {R.}~\bibnamefont {Kolesov}}, \bibinfo {author} {\bibfnamefont {N.}~\bibnamefont {Mizuochi}}, \bibinfo {author} {\bibfnamefont {J.}~\bibnamefont {Isoya}}, \bibinfo {author} {\bibfnamefont {J.}~\bibnamefont {Achard}}, \bibinfo {author} {\bibfnamefont {J.}~\bibnamefont {Beck}}, \bibinfo {author} {\bibfnamefont {J.}~\bibnamefont {Tissler}}, \bibinfo {author} {\bibfnamefont {V.}~\bibnamefont {Jacques}}, \bibinfo {author} {\bibfnamefont {P.~R.}\ \bibnamefont {Hemmer}}, \bibinfo {author} {\bibfnamefont {F.}~\bibnamefont {Jelezko}}, \ and\ \bibinfo {author} {\bibfnamefont {J.}~\bibnamefont {Wrachtrup}},\ }\href {\doibase 10.1038/nmat2420} {\bibfield  {journal} {\bibinfo  {journal} {Nat. Mater.}\ }\textbf
  {\bibinfo {volume} {8}},\ \bibinfo {pages} {383 – 387} (\bibinfo {year} {2009})}\BibitemShut {NoStop}%
\bibitem [{\citenamefont {de~Lange}\ \emph {et~al.}(2010)\citenamefont {de~Lange}, \citenamefont {Wang}, \citenamefont {Ristè}, \citenamefont {Dobrovitski},\ and\ \citenamefont {Hanson}}]{science_deLange_1}%
  \BibitemOpen
  \bibfield  {author} {\bibinfo {author} {\bibfnamefont {G.}~\bibnamefont {de~Lange}}, \bibinfo {author} {\bibfnamefont {Z.~H.}\ \bibnamefont {Wang}}, \bibinfo {author} {\bibfnamefont {D.}~\bibnamefont {Ristè}}, \bibinfo {author} {\bibfnamefont {V.~V.}\ \bibnamefont {Dobrovitski}}, \ and\ \bibinfo {author} {\bibfnamefont {R.}~\bibnamefont {Hanson}},\ }\href {\doibase 10.1126/science.1192739} {\bibfield  {journal} {\bibinfo  {journal} {Science}\ }\textbf {\bibinfo {volume} {330}},\ \bibinfo {pages} {60} (\bibinfo {year} {2010})}\BibitemShut {NoStop}%
\bibitem [{\citenamefont {Degen}(2008)}]{APL_Degen}%
  \BibitemOpen
  \bibfield  {author} {\bibinfo {author} {\bibfnamefont {C.~L.}\ \bibnamefont {Degen}},\ }\href {\doibase 10.1063/1.2943282} {\bibfield  {journal} {\bibinfo  {journal} {Appl. Phys. Lett.}\ }\textbf {\bibinfo {volume} {92}},\ \bibinfo {pages} {243111} (\bibinfo {year} {2008})}\BibitemShut {NoStop}%
\bibitem [{\citenamefont {Balasubramanian}\ \emph {et~al.}(2008)\citenamefont {Balasubramanian}, \citenamefont {Chan}, \citenamefont {Kolesov}, \citenamefont {Al-Hmoud}, \citenamefont {Tisler}, \citenamefont {Shin}, \citenamefont {Kim}, \citenamefont {Wojcik}, \citenamefont {Hemmer}, \citenamefont {Krueger}, \citenamefont {Hanke}, \citenamefont {Leitenstorfer}, \citenamefont {Bratschitsch}, \citenamefont {Jelezko},\ and\ \citenamefont {Wrachtrup}}]{Nature_Balasubramanian_1}%
  \BibitemOpen
  \bibfield  {author} {\bibinfo {author} {\bibfnamefont {G.}~\bibnamefont {Balasubramanian}}, \bibinfo {author} {\bibfnamefont {I.~Y.}\ \bibnamefont {Chan}}, \bibinfo {author} {\bibfnamefont {R.}~\bibnamefont {Kolesov}}, \bibinfo {author} {\bibfnamefont {M.}~\bibnamefont {Al-Hmoud}}, \bibinfo {author} {\bibfnamefont {J.}~\bibnamefont {Tisler}}, \bibinfo {author} {\bibfnamefont {C.}~\bibnamefont {Shin}}, \bibinfo {author} {\bibfnamefont {C.}~\bibnamefont {Kim}}, \bibinfo {author} {\bibfnamefont {A.}~\bibnamefont {Wojcik}}, \bibinfo {author} {\bibfnamefont {P.~R.}\ \bibnamefont {Hemmer}}, \bibinfo {author} {\bibfnamefont {A.}~\bibnamefont {Krueger}}, \bibinfo {author} {\bibfnamefont {T.}~\bibnamefont {Hanke}}, \bibinfo {author} {\bibfnamefont {A.}~\bibnamefont {Leitenstorfer}}, \bibinfo {author} {\bibfnamefont {R.}~\bibnamefont {Bratschitsch}}, \bibinfo {author} {\bibfnamefont {F.}~\bibnamefont {Jelezko}}, \ and\ \bibinfo {author} {\bibfnamefont {J.}~\bibnamefont {Wrachtrup}},\ }\href {\doibase
  10.1038/nature07278} {\bibfield  {journal} {\bibinfo  {journal} {Nature}\ }\textbf {\bibinfo {volume} {455}},\ \bibinfo {pages} {648} (\bibinfo {year} {2008})}\BibitemShut {NoStop}%
\bibitem [{\citenamefont {Maze}\ \emph {et~al.}(2008)\citenamefont {Maze}, \citenamefont {Stanwix}, \citenamefont {Hodges}, \citenamefont {Hong}, \citenamefont {Taylor}, \citenamefont {Cappellaro}, \citenamefont {Jiang}, \citenamefont {Dutt}, \citenamefont {Togan}, \citenamefont {Zibrov}, \citenamefont {Yacoby}, \citenamefont {Walsworth},\ and\ \citenamefont {Lukin}}]{Nature_Maze_1}%
  \BibitemOpen
  \bibfield  {author} {\bibinfo {author} {\bibfnamefont {J.~R.}\ \bibnamefont {Maze}}, \bibinfo {author} {\bibfnamefont {P.~L.}\ \bibnamefont {Stanwix}}, \bibinfo {author} {\bibfnamefont {J.~S.}\ \bibnamefont {Hodges}}, \bibinfo {author} {\bibfnamefont {S.}~\bibnamefont {Hong}}, \bibinfo {author} {\bibfnamefont {J.~M.}\ \bibnamefont {Taylor}}, \bibinfo {author} {\bibfnamefont {P.}~\bibnamefont {Cappellaro}}, \bibinfo {author} {\bibfnamefont {L.}~\bibnamefont {Jiang}}, \bibinfo {author} {\bibfnamefont {M.~V.~G.}\ \bibnamefont {Dutt}}, \bibinfo {author} {\bibfnamefont {E.}~\bibnamefont {Togan}}, \bibinfo {author} {\bibfnamefont {A.~S.}\ \bibnamefont {Zibrov}}, \bibinfo {author} {\bibfnamefont {A.}~\bibnamefont {Yacoby}}, \bibinfo {author} {\bibfnamefont {R.~L.}\ \bibnamefont {Walsworth}}, \ and\ \bibinfo {author} {\bibfnamefont {M.~D.}\ \bibnamefont {Lukin}},\ }\href {\doibase 10.1038/nature07279} {\bibfield  {journal} {\bibinfo  {journal} {Nature}\ }\textbf {\bibinfo {volume} {455}},\ \bibinfo {pages} {644}
  (\bibinfo {year} {2008})}\BibitemShut {NoStop}%
\bibitem [{\citenamefont {Taylor}\ \emph {et~al.}(2008)\citenamefont {Taylor}, \citenamefont {Cappellaro}, \citenamefont {Childress}, \citenamefont {Jiang}, \citenamefont {Budker}, \citenamefont {Hemmer}, \citenamefont {Yacoby}, \citenamefont {Walsworth},\ and\ \citenamefont {Lukin}}]{Nature_Taylor_1}%
  \BibitemOpen
  \bibfield  {author} {\bibinfo {author} {\bibfnamefont {J.~M.}\ \bibnamefont {Taylor}}, \bibinfo {author} {\bibfnamefont {P.}~\bibnamefont {Cappellaro}}, \bibinfo {author} {\bibfnamefont {L.}~\bibnamefont {Childress}}, \bibinfo {author} {\bibfnamefont {L.}~\bibnamefont {Jiang}}, \bibinfo {author} {\bibfnamefont {D.}~\bibnamefont {Budker}}, \bibinfo {author} {\bibfnamefont {P.~R.}\ \bibnamefont {Hemmer}}, \bibinfo {author} {\bibfnamefont {A.}~\bibnamefont {Yacoby}}, \bibinfo {author} {\bibfnamefont {R.}~\bibnamefont {Walsworth}}, \ and\ \bibinfo {author} {\bibfnamefont {M.~D.}\ \bibnamefont {Lukin}},\ }\href {\doibase 10.1038/nphys1075} {\bibfield  {journal} {\bibinfo  {journal} {Nat. Phys.}\ }\textbf {\bibinfo {volume} {4}},\ \bibinfo {pages} {810} (\bibinfo {year} {2008})}\BibitemShut {NoStop}%
\bibitem [{\citenamefont {Grinolds}\ \emph {et~al.}(2013)\citenamefont {Grinolds}, \citenamefont {Hong}, \citenamefont {Maletinsky}, \citenamefont {Luan}, \citenamefont {Lukin}, \citenamefont {Walsworth},\ and\ \citenamefont {Yacoby}}]{Nature_Grinolds_2}%
  \BibitemOpen
  \bibfield  {author} {\bibinfo {author} {\bibfnamefont {M.~S.}\ \bibnamefont {Grinolds}}, \bibinfo {author} {\bibfnamefont {S.}~\bibnamefont {Hong}}, \bibinfo {author} {\bibfnamefont {P.}~\bibnamefont {Maletinsky}}, \bibinfo {author} {\bibfnamefont {L.}~\bibnamefont {Luan}}, \bibinfo {author} {\bibfnamefont {M.~D.}\ \bibnamefont {Lukin}}, \bibinfo {author} {\bibfnamefont {R.~L.}\ \bibnamefont {Walsworth}}, \ and\ \bibinfo {author} {\bibfnamefont {A.}~\bibnamefont {Yacoby}},\ }\href {\doibase 10.1038/NPHYS2543} {\bibfield  {journal} {\bibinfo  {journal} {Nat. Phys.}\ }\textbf {\bibinfo {volume} {9}},\ \bibinfo {pages} {215} (\bibinfo {year} {2013})}\BibitemShut {NoStop}%
\bibitem [{\citenamefont {Shi}\ \emph {et~al.}(2015)\citenamefont {Shi}, \citenamefont {Zhang}, \citenamefont {Wang}, \citenamefont {Sun}, \citenamefont {Wang}, \citenamefont {Rong}, \citenamefont {Chen}, \citenamefont {Ju}, \citenamefont {Reinhard}, \citenamefont {Chen}, \citenamefont {Wrachtrup}, \citenamefont {Wang},\ and\ \citenamefont {Du}}]{Science_Shi_2}%
  \BibitemOpen
  \bibfield  {author} {\bibinfo {author} {\bibfnamefont {F.}~\bibnamefont {Shi}}, \bibinfo {author} {\bibfnamefont {Q.}~\bibnamefont {Zhang}}, \bibinfo {author} {\bibfnamefont {P.}~\bibnamefont {Wang}}, \bibinfo {author} {\bibfnamefont {H.}~\bibnamefont {Sun}}, \bibinfo {author} {\bibfnamefont {J.}~\bibnamefont {Wang}}, \bibinfo {author} {\bibfnamefont {X.}~\bibnamefont {Rong}}, \bibinfo {author} {\bibfnamefont {M.}~\bibnamefont {Chen}}, \bibinfo {author} {\bibfnamefont {C.}~\bibnamefont {Ju}}, \bibinfo {author} {\bibfnamefont {F.}~\bibnamefont {Reinhard}}, \bibinfo {author} {\bibfnamefont {H.}~\bibnamefont {Chen}}, \bibinfo {author} {\bibfnamefont {J.}~\bibnamefont {Wrachtrup}}, \bibinfo {author} {\bibfnamefont {J.}~\bibnamefont {Wang}}, \ and\ \bibinfo {author} {\bibfnamefont {J.}~\bibnamefont {Du}},\ }\href {\doibase 10.1126/science.aaa2253} {\bibfield  {journal} {\bibinfo  {journal} {Science}\ }\textbf {\bibinfo {volume} {347}},\ \bibinfo {pages} {1135} (\bibinfo {year} {2015})}\BibitemShut {NoStop}%
\bibitem [{\citenamefont {Abeywardana}\ \emph {et~al.}(2016)\citenamefont {Abeywardana}, \citenamefont {Stepanov}, \citenamefont {Cho},\ and\ \citenamefont {Takahashi}}]{JAP_Abeywardana_1}%
  \BibitemOpen
  \bibfield  {author} {\bibinfo {author} {\bibfnamefont {C.}~\bibnamefont {Abeywardana}}, \bibinfo {author} {\bibfnamefont {V.}~\bibnamefont {Stepanov}}, \bibinfo {author} {\bibfnamefont {F.~H.}\ \bibnamefont {Cho}}, \ and\ \bibinfo {author} {\bibfnamefont {S.}~\bibnamefont {Takahashi}},\ }\href {\doibase 10.1063/1.4963717} {\bibfield  {journal} {\bibinfo  {journal} {J. Appl. Phys.}\ }\textbf {\bibinfo {volume} {120}} (\bibinfo {year} {2016}),\ 10.1063/1.4963717}\BibitemShut {NoStop}%
\bibitem [{\citenamefont {Fortman}\ \emph {et~al.}(2021)\citenamefont {Fortman}, \citenamefont {Mugica-Sanchez}, \citenamefont {Tischler}, \citenamefont {Selco}, \citenamefont {Hang}, \citenamefont {Holczer},\ and\ \citenamefont {Takahashi}}]{JAP_Fortman_1}%
  \BibitemOpen
  \bibfield  {author} {\bibinfo {author} {\bibfnamefont {B.}~\bibnamefont {Fortman}}, \bibinfo {author} {\bibfnamefont {L.}~\bibnamefont {Mugica-Sanchez}}, \bibinfo {author} {\bibfnamefont {N.}~\bibnamefont {Tischler}}, \bibinfo {author} {\bibfnamefont {C.}~\bibnamefont {Selco}}, \bibinfo {author} {\bibfnamefont {Y.}~\bibnamefont {Hang}}, \bibinfo {author} {\bibfnamefont {K.}~\bibnamefont {Holczer}}, \ and\ \bibinfo {author} {\bibfnamefont {S.}~\bibnamefont {Takahashi}},\ }\href {\doibase 10.1063/5.0055642} {\bibfield  {journal} {\bibinfo  {journal} {J. Appl. Phys.}\ }\textbf {\bibinfo {volume} {130}},\ \bibinfo {pages} {083901} (\bibinfo {year} {2021})}\BibitemShut {NoStop}%
\bibitem [{\citenamefont {Fortman}\ and\ \citenamefont {Takahashi}(2019)}]{JPC_Fortman_1}%
  \BibitemOpen
  \bibfield  {author} {\bibinfo {author} {\bibfnamefont {B.}~\bibnamefont {Fortman}}\ and\ \bibinfo {author} {\bibfnamefont {S.}~\bibnamefont {Takahashi}},\ }\href {\doibase 10.1021/acs.jpca.9b02445} {\bibfield  {journal} {\bibinfo  {journal} {J. Phys. Chem. A}\ }\textbf {\bibinfo {volume} {123}},\ \bibinfo {pages} {6350} (\bibinfo {year} {2019})}\BibitemShut {NoStop}%
\bibitem [{\citenamefont {Fortman}\ \emph {et~al.}(2020)\citenamefont {Fortman}, \citenamefont {Pena}, \citenamefont {Holczer},\ and\ \citenamefont {Takahashi}}]{APL_Fortman_1}%
  \BibitemOpen
  \bibfield  {author} {\bibinfo {author} {\bibfnamefont {B.}~\bibnamefont {Fortman}}, \bibinfo {author} {\bibfnamefont {J.}~\bibnamefont {Pena}}, \bibinfo {author} {\bibfnamefont {K.}~\bibnamefont {Holczer}}, \ and\ \bibinfo {author} {\bibfnamefont {S.}~\bibnamefont {Takahashi}},\ }\href {\doibase 10.1063/5.0006014} {\bibfield  {journal} {\bibinfo  {journal} {Appl. Phys. Lett.}\ }\textbf {\bibinfo {volume} {116}},\ \bibinfo {pages} {174004} (\bibinfo {year} {2020})}\BibitemShut {NoStop}%
\bibitem [{\citenamefont {Li}\ \emph {et~al.}(2021)\citenamefont {Li}, \citenamefont {Zheng}, \citenamefont {Peng}, \citenamefont {Kamiya}, \citenamefont {Niki}, \citenamefont {Stepanov}, \citenamefont {Jarmola}, \citenamefont {Shimizu}, \citenamefont {Takahashi}, \citenamefont {Wickenbrock},\ and\ \citenamefont {Budker}}]{PRB_LZP_1}%
  \BibitemOpen
  \bibfield  {author} {\bibinfo {author} {\bibfnamefont {S.}~\bibnamefont {Li}}, \bibinfo {author} {\bibfnamefont {H.}~\bibnamefont {Zheng}}, \bibinfo {author} {\bibfnamefont {Z.}~\bibnamefont {Peng}}, \bibinfo {author} {\bibfnamefont {M.}~\bibnamefont {Kamiya}}, \bibinfo {author} {\bibfnamefont {T.}~\bibnamefont {Niki}}, \bibinfo {author} {\bibfnamefont {V.}~\bibnamefont {Stepanov}}, \bibinfo {author} {\bibfnamefont {A.}~\bibnamefont {Jarmola}}, \bibinfo {author} {\bibfnamefont {Y.}~\bibnamefont {Shimizu}}, \bibinfo {author} {\bibfnamefont {S.}~\bibnamefont {Takahashi}}, \bibinfo {author} {\bibfnamefont {A.}~\bibnamefont {Wickenbrock}}, \ and\ \bibinfo {author} {\bibfnamefont {D.}~\bibnamefont {Budker}},\ }\href {\doibase 10.1103/PhysRevB.104.094307} {\bibfield  {journal} {\bibinfo  {journal} {Phys. Rev. B}\ }\textbf {\bibinfo {volume} {104}},\ \bibinfo {pages} {094307} (\bibinfo {year} {2021})}\BibitemShut {NoStop}%
\bibitem [{\citenamefont {Ren}\ \emph {et~al.}(2023)\citenamefont {Ren}, \citenamefont {Selco}, \citenamefont {Kawashiri}, \citenamefont {Coumans}, \citenamefont {Fortman}, \citenamefont {Bouchard}, \citenamefont {Holczer},\ and\ \citenamefont {Takahashi}}]{PRB_Ren_1}%
  \BibitemOpen
  \bibfield  {author} {\bibinfo {author} {\bibfnamefont {Y.}~\bibnamefont {Ren}}, \bibinfo {author} {\bibfnamefont {C.}~\bibnamefont {Selco}}, \bibinfo {author} {\bibfnamefont {D.}~\bibnamefont {Kawashiri}}, \bibinfo {author} {\bibfnamefont {M.}~\bibnamefont {Coumans}}, \bibinfo {author} {\bibfnamefont {B.}~\bibnamefont {Fortman}}, \bibinfo {author} {\bibfnamefont {L.-S.}\ \bibnamefont {Bouchard}}, \bibinfo {author} {\bibfnamefont {K.}~\bibnamefont {Holczer}}, \ and\ \bibinfo {author} {\bibfnamefont {S.}~\bibnamefont {Takahashi}},\ }\href {\doibase 10.1103/PhysRevB.108.045421} {\bibfield  {journal} {\bibinfo  {journal} {Phys. Rev. B}\ }\textbf {\bibinfo {volume} {108}},\ \bibinfo {pages} {045421} (\bibinfo {year} {2023})}\BibitemShut {NoStop}%
\bibitem [{\citenamefont {Cai}\ \emph {et~al.}(2013)\citenamefont {Cai}, \citenamefont {Retzker}, \citenamefont {Jelezko},\ and\ \citenamefont {Plenio}}]{Nature_Cai_1}%
  \BibitemOpen
  \bibfield  {author} {\bibinfo {author} {\bibfnamefont {J.}~\bibnamefont {Cai}}, \bibinfo {author} {\bibfnamefont {A.}~\bibnamefont {Retzker}}, \bibinfo {author} {\bibfnamefont {F.}~\bibnamefont {Jelezko}}, \ and\ \bibinfo {author} {\bibfnamefont {M.~B.}\ \bibnamefont {Plenio}},\ }\href {\doibase 10.1038/nphys2519} {\bibfield  {journal} {\bibinfo  {journal} {Nat. Phys.}\ }\textbf {\bibinfo {volume} {9}},\ \bibinfo {pages} {168} (\bibinfo {year} {2013})}\BibitemShut {NoStop}%
\bibitem [{\citenamefont {Wang}\ \emph {et~al.}(2015)\citenamefont {Wang}, \citenamefont {Dolde}, \citenamefont {Biamonte}, \citenamefont {Babbush}, \citenamefont {Bergholm}, \citenamefont {Yang}, \citenamefont {Jakobi}, \citenamefont {Neumann}, \citenamefont {Aspuru-Guzik}, \citenamefont {Whitfield},\ and\ \citenamefont {Wrachtrup}}]{ACS_Wang_1}%
  \BibitemOpen
  \bibfield  {author} {\bibinfo {author} {\bibfnamefont {Y.}~\bibnamefont {Wang}}, \bibinfo {author} {\bibfnamefont {F.}~\bibnamefont {Dolde}}, \bibinfo {author} {\bibfnamefont {J.}~\bibnamefont {Biamonte}}, \bibinfo {author} {\bibfnamefont {R.}~\bibnamefont {Babbush}}, \bibinfo {author} {\bibfnamefont {V.}~\bibnamefont {Bergholm}}, \bibinfo {author} {\bibfnamefont {S.}~\bibnamefont {Yang}}, \bibinfo {author} {\bibfnamefont {I.}~\bibnamefont {Jakobi}}, \bibinfo {author} {\bibfnamefont {P.}~\bibnamefont {Neumann}}, \bibinfo {author} {\bibfnamefont {A.}~\bibnamefont {Aspuru-Guzik}}, \bibinfo {author} {\bibfnamefont {J.~D.}\ \bibnamefont {Whitfield}}, \ and\ \bibinfo {author} {\bibfnamefont {J.}~\bibnamefont {Wrachtrup}},\ }\href {\doibase 10.1021/acsnano.5b01651} {\bibfield  {journal} {\bibinfo  {journal} {ACS Nano}\ }\textbf {\bibinfo {volume} {9}},\ \bibinfo {pages} {7769} (\bibinfo {year} {2015})}\BibitemShut {NoStop}%
\bibitem [{\citenamefont {Ju}\ \emph {et~al.}(2014)\citenamefont {Ju}, \citenamefont {Lei}, \citenamefont {Xu}, \citenamefont {Culcer}, \citenamefont {Zhang},\ and\ \citenamefont {Du}}]{PRB_Ju_1}%
  \BibitemOpen
  \bibfield  {author} {\bibinfo {author} {\bibfnamefont {C.}~\bibnamefont {Ju}}, \bibinfo {author} {\bibfnamefont {C.}~\bibnamefont {Lei}}, \bibinfo {author} {\bibfnamefont {X.}~\bibnamefont {Xu}}, \bibinfo {author} {\bibfnamefont {D.}~\bibnamefont {Culcer}}, \bibinfo {author} {\bibfnamefont {Z.}~\bibnamefont {Zhang}}, \ and\ \bibinfo {author} {\bibfnamefont {J.}~\bibnamefont {Du}},\ }\href {\doibase 10.1103/PhysRevB.89.045432} {\bibfield  {journal} {\bibinfo  {journal} {Phys. Rev. B}\ }\textbf {\bibinfo {volume} {89}},\ \bibinfo {pages} {045432} (\bibinfo {year} {2014})}\BibitemShut {NoStop}%
\bibitem [{\citenamefont {Xie}\ \emph {et~al.}(2021)\citenamefont {Xie}, \citenamefont {Zhao}, \citenamefont {Kong}, \citenamefont {Ma}, \citenamefont {Wang}, \citenamefont {Ye}, \citenamefont {Yu}, \citenamefont {Yang}, \citenamefont {Xu}, \citenamefont {Wang}, \citenamefont {Wang}, \citenamefont {Shi},\ and\ \citenamefont {Du}}]{Science_Xie_1}%
  \BibitemOpen
  \bibfield  {author} {\bibinfo {author} {\bibfnamefont {T.}~\bibnamefont {Xie}}, \bibinfo {author} {\bibfnamefont {Z.}~\bibnamefont {Zhao}}, \bibinfo {author} {\bibfnamefont {X.}~\bibnamefont {Kong}}, \bibinfo {author} {\bibfnamefont {W.}~\bibnamefont {Ma}}, \bibinfo {author} {\bibfnamefont {M.}~\bibnamefont {Wang}}, \bibinfo {author} {\bibfnamefont {X.}~\bibnamefont {Ye}}, \bibinfo {author} {\bibfnamefont {P.}~\bibnamefont {Yu}}, \bibinfo {author} {\bibfnamefont {Z.}~\bibnamefont {Yang}}, \bibinfo {author} {\bibfnamefont {S.}~\bibnamefont {Xu}}, \bibinfo {author} {\bibfnamefont {P.}~\bibnamefont {Wang}}, \bibinfo {author} {\bibfnamefont {Y.}~\bibnamefont {Wang}}, \bibinfo {author} {\bibfnamefont {F.}~\bibnamefont {Shi}}, \ and\ \bibinfo {author} {\bibfnamefont {J.}~\bibnamefont {Du}},\ }\href {\doibase 10.1126/sciadv.abg9204} {\bibfield  {journal} {\bibinfo  {journal} {Sci. Adv.}\ }\textbf {\bibinfo {volume} {7}},\ \bibinfo {pages} {eabg9204} (\bibinfo {year} {2021})}\BibitemShut {NoStop}%
\bibitem [{\citenamefont {Broholm}\ \emph {et~al.}(2020)\citenamefont {Broholm}, \citenamefont {Cava}, \citenamefont {Kivelson}, \citenamefont {Nocera}, \citenamefont {Norman},\ and\ \citenamefont {Senthil}}]{Science_Broholm}%
  \BibitemOpen
  \bibfield  {author} {\bibinfo {author} {\bibfnamefont {C.}~\bibnamefont {Broholm}}, \bibinfo {author} {\bibfnamefont {R.~J.}\ \bibnamefont {Cava}}, \bibinfo {author} {\bibfnamefont {S.~A.}\ \bibnamefont {Kivelson}}, \bibinfo {author} {\bibfnamefont {D.~G.}\ \bibnamefont {Nocera}}, \bibinfo {author} {\bibfnamefont {M.~R.}\ \bibnamefont {Norman}}, \ and\ \bibinfo {author} {\bibfnamefont {T.}~\bibnamefont {Senthil}},\ }\href {\doibase 10.1126/science.aay0668} {\bibfield  {journal} {\bibinfo  {journal} {Science}\ }\textbf {\bibinfo {volume} {367}},\ \bibinfo {pages} {eaay0668} (\bibinfo {year} {2020})}\BibitemShut {NoStop}%
\bibitem [{\citenamefont {Choi}\ \emph {et~al.}(2017)\citenamefont {Choi}, \citenamefont {Choi}, \citenamefont {Landig}, \citenamefont {Kucsko}, \citenamefont {Zhou}, \citenamefont {Isoya}, \citenamefont {Jelezko}, \citenamefont {Onoda}, \citenamefont {Sumiya}, \citenamefont {Khemani}, \citenamefont {von Keyserlingk}, \citenamefont {Yao}, \citenamefont {Demler},\ and\ \citenamefont {Lukin}}]{Nature_Choi_1}%
  \BibitemOpen
  \bibfield  {author} {\bibinfo {author} {\bibfnamefont {S.}~\bibnamefont {Choi}}, \bibinfo {author} {\bibfnamefont {J.}~\bibnamefont {Choi}}, \bibinfo {author} {\bibfnamefont {R.}~\bibnamefont {Landig}}, \bibinfo {author} {\bibfnamefont {G.}~\bibnamefont {Kucsko}}, \bibinfo {author} {\bibfnamefont {H.}~\bibnamefont {Zhou}}, \bibinfo {author} {\bibfnamefont {J.}~\bibnamefont {Isoya}}, \bibinfo {author} {\bibfnamefont {F.}~\bibnamefont {Jelezko}}, \bibinfo {author} {\bibfnamefont {S.}~\bibnamefont {Onoda}}, \bibinfo {author} {\bibfnamefont {H.}~\bibnamefont {Sumiya}}, \bibinfo {author} {\bibfnamefont {V.}~\bibnamefont {Khemani}}, \bibinfo {author} {\bibfnamefont {C.}~\bibnamefont {von Keyserlingk}}, \bibinfo {author} {\bibfnamefont {N.~Y.}\ \bibnamefont {Yao}}, \bibinfo {author} {\bibfnamefont {E.}~\bibnamefont {Demler}}, \ and\ \bibinfo {author} {\bibfnamefont {M.~D.}\ \bibnamefont {Lukin}},\ }\href {\doibase 10.1038/nature21426} {\bibfield  {journal} {\bibinfo  {journal} {Nature}\ }\textbf {\bibinfo {volume}
  {543}},\ \bibinfo {pages} {221+} (\bibinfo {year} {2017})}\BibitemShut {NoStop}%
\bibitem [{\citenamefont {Gonz\'alez}\ \emph {et~al.}(2022)\citenamefont {Gonz\'alez}, \citenamefont {Norambuena},\ and\ \citenamefont {Coto}}]{PRB_Francisco}%
  \BibitemOpen
  \bibfield  {author} {\bibinfo {author} {\bibfnamefont {F.~J.}\ \bibnamefont {Gonz\'alez}}, \bibinfo {author} {\bibfnamefont {A.}~\bibnamefont {Norambuena}}, \ and\ \bibinfo {author} {\bibfnamefont {R.}~\bibnamefont {Coto}},\ }\href {\doibase 10.1103/PhysRevB.106.014313} {\bibfield  {journal} {\bibinfo  {journal} {Phys. Rev. B}\ }\textbf {\bibinfo {volume} {106}},\ \bibinfo {pages} {014313} (\bibinfo {year} {2022})}\BibitemShut {NoStop}%
\bibitem [{\citenamefont {Bussandri}\ \emph {et~al.}(2024)\citenamefont {Bussandri}, \citenamefont {Shimon}, \citenamefont {Equbal}, \citenamefont {Ren}, \citenamefont {Takahashi}, \citenamefont {Ramanathan},\ and\ \citenamefont {Han}}]{JACS_Bussandri_1}%
  \BibitemOpen
  \bibfield  {author} {\bibinfo {author} {\bibfnamefont {S.}~\bibnamefont {Bussandri}}, \bibinfo {author} {\bibfnamefont {D.}~\bibnamefont {Shimon}}, \bibinfo {author} {\bibfnamefont {A.}~\bibnamefont {Equbal}}, \bibinfo {author} {\bibfnamefont {Y.}~\bibnamefont {Ren}}, \bibinfo {author} {\bibfnamefont {S.}~\bibnamefont {Takahashi}}, \bibinfo {author} {\bibfnamefont {C.}~\bibnamefont {Ramanathan}}, \ and\ \bibinfo {author} {\bibfnamefont {S.}~\bibnamefont {Han}},\ }\href {\doibase 10.1021/jacs.3c06705} {\bibfield  {journal} {\bibinfo  {journal} {J. Am. Chem. Soc.}\ }\textbf {\bibinfo {volume} {146}},\ \bibinfo {pages} {5088} (\bibinfo {year} {2024})}\BibitemShut {NoStop}%
\bibitem [{\citenamefont {Yamamoto}\ \emph {et~al.}(2013)\citenamefont {Yamamoto}, \citenamefont {M\"uller}, \citenamefont {McGuinness}, \citenamefont {Teraji}, \citenamefont {Naydenov}, \citenamefont {Onoda}, \citenamefont {Ohshima}, \citenamefont {Wrachtrup}, \citenamefont {Jelezko},\ and\ \citenamefont {Isoya}}]{PRB_Yamamoto_1}%
  \BibitemOpen
  \bibfield  {author} {\bibinfo {author} {\bibfnamefont {T.}~\bibnamefont {Yamamoto}}, \bibinfo {author} {\bibfnamefont {C.}~\bibnamefont {M\"uller}}, \bibinfo {author} {\bibfnamefont {L.~P.}\ \bibnamefont {McGuinness}}, \bibinfo {author} {\bibfnamefont {T.}~\bibnamefont {Teraji}}, \bibinfo {author} {\bibfnamefont {B.}~\bibnamefont {Naydenov}}, \bibinfo {author} {\bibfnamefont {S.}~\bibnamefont {Onoda}}, \bibinfo {author} {\bibfnamefont {T.}~\bibnamefont {Ohshima}}, \bibinfo {author} {\bibfnamefont {J.}~\bibnamefont {Wrachtrup}}, \bibinfo {author} {\bibfnamefont {F.}~\bibnamefont {Jelezko}}, \ and\ \bibinfo {author} {\bibfnamefont {J.}~\bibnamefont {Isoya}},\ }\href {\doibase 10.1103/PhysRevB.88.201201} {\bibfield  {journal} {\bibinfo  {journal} {Phys. Rev. B}\ }\textbf {\bibinfo {volume} {88}},\ \bibinfo {pages} {201201} (\bibinfo {year} {2013})}\BibitemShut {NoStop}%
\bibitem [{\citenamefont {Shi}\ \emph {et~al.}(2013)\citenamefont {Shi}, \citenamefont {Zhang}, \citenamefont {Naydenov}, \citenamefont {Jelezko}, \citenamefont {Du}, \citenamefont {Reinhard},\ and\ \citenamefont {Wrachtrup}}]{PRB_Shi_1}%
  \BibitemOpen
  \bibfield  {author} {\bibinfo {author} {\bibfnamefont {F.}~\bibnamefont {Shi}}, \bibinfo {author} {\bibfnamefont {Q.}~\bibnamefont {Zhang}}, \bibinfo {author} {\bibfnamefont {B.}~\bibnamefont {Naydenov}}, \bibinfo {author} {\bibfnamefont {F.}~\bibnamefont {Jelezko}}, \bibinfo {author} {\bibfnamefont {J.}~\bibnamefont {Du}}, \bibinfo {author} {\bibfnamefont {F.}~\bibnamefont {Reinhard}}, \ and\ \bibinfo {author} {\bibfnamefont {J.}~\bibnamefont {Wrachtrup}},\ }\href {\doibase 10.1103/PhysRevB.87.195414} {\bibfield  {journal} {\bibinfo  {journal} {Phys. Rev. B}\ }\textbf {\bibinfo {volume} {87}},\ \bibinfo {pages} {195414} (\bibinfo {year} {2013})}\BibitemShut {NoStop}%
\bibitem [{\citenamefont {Cooper}\ \emph {et~al.}(2020)\citenamefont {Cooper}, \citenamefont {Sun}, \citenamefont {Jaskula},\ and\ \citenamefont {Cappellaro}}]{PRL_Cooper_1}%
  \BibitemOpen
  \bibfield  {author} {\bibinfo {author} {\bibfnamefont {A.}~\bibnamefont {Cooper}}, \bibinfo {author} {\bibfnamefont {W.~K.~C.}\ \bibnamefont {Sun}}, \bibinfo {author} {\bibfnamefont {J.-C.}\ \bibnamefont {Jaskula}}, \ and\ \bibinfo {author} {\bibfnamefont {P.}~\bibnamefont {Cappellaro}},\ }\href {\doibase 10.1103/PhysRevLett.124.083602} {\bibfield  {journal} {\bibinfo  {journal} {Phys. Rev. Lett.}\ }\textbf {\bibinfo {volume} {124}},\ \bibinfo {pages} {083602} (\bibinfo {year} {2020})}\BibitemShut {NoStop}%
\bibitem [{\citenamefont {Rosenfeld}\ \emph {et~al.}(2018)\citenamefont {Rosenfeld}, \citenamefont {Pham}, \citenamefont {Lukin},\ and\ \citenamefont {Walsworth}}]{PRL_Rosenfeld_1}%
  \BibitemOpen
  \bibfield  {author} {\bibinfo {author} {\bibfnamefont {E.~L.}\ \bibnamefont {Rosenfeld}}, \bibinfo {author} {\bibfnamefont {L.~M.}\ \bibnamefont {Pham}}, \bibinfo {author} {\bibfnamefont {M.~D.}\ \bibnamefont {Lukin}}, \ and\ \bibinfo {author} {\bibfnamefont {R.~L.}\ \bibnamefont {Walsworth}},\ }\href {\doibase 10.1103/PhysRevLett.120.243604} {\bibfield  {journal} {\bibinfo  {journal} {Phys. Rev. Lett.}\ }\textbf {\bibinfo {volume} {120}},\ \bibinfo {pages} {243604} (\bibinfo {year} {2018})}\BibitemShut {NoStop}%
\bibitem [{\citenamefont {Grinolds}\ \emph {et~al.}(2014)\citenamefont {Grinolds}, \citenamefont {Warner}, \citenamefont {De~Greve}, \citenamefont {Dovzhenko}, \citenamefont {Thiel}, \citenamefont {Walsworth}, \citenamefont {Hong}, \citenamefont {Maletinsky},\ and\ \citenamefont {Yacoby}}]{nature_Grinolds_1}%
  \BibitemOpen
  \bibfield  {author} {\bibinfo {author} {\bibfnamefont {M.~S.}\ \bibnamefont {Grinolds}}, \bibinfo {author} {\bibfnamefont {M.}~\bibnamefont {Warner}}, \bibinfo {author} {\bibfnamefont {K.}~\bibnamefont {De~Greve}}, \bibinfo {author} {\bibfnamefont {Y.}~\bibnamefont {Dovzhenko}}, \bibinfo {author} {\bibfnamefont {L.}~\bibnamefont {Thiel}}, \bibinfo {author} {\bibfnamefont {R.~L.}\ \bibnamefont {Walsworth}}, \bibinfo {author} {\bibfnamefont {S.}~\bibnamefont {Hong}}, \bibinfo {author} {\bibfnamefont {P.}~\bibnamefont {Maletinsky}}, \ and\ \bibinfo {author} {\bibfnamefont {A.}~\bibnamefont {Yacoby}},\ }\href {\doibase 10.1038/nnano.2014.30} {\bibfield  {journal} {\bibinfo  {journal} {Nat. Nanotechnol.}\ }\textbf {\bibinfo {volume} {9}},\ \bibinfo {pages} {279} (\bibinfo {year} {2014})}\BibitemShut {NoStop}%
\bibitem [{\citenamefont {Sushkov}\ \emph {et~al.}(2014)\citenamefont {Sushkov}, \citenamefont {Lovchinsky}, \citenamefont {Chisholm}, \citenamefont {Walsworth}, \citenamefont {Park},\ and\ \citenamefont {Lukin}}]{PRL_Sushkov_1}%
  \BibitemOpen
  \bibfield  {author} {\bibinfo {author} {\bibfnamefont {A.~O.}\ \bibnamefont {Sushkov}}, \bibinfo {author} {\bibfnamefont {I.}~\bibnamefont {Lovchinsky}}, \bibinfo {author} {\bibfnamefont {N.}~\bibnamefont {Chisholm}}, \bibinfo {author} {\bibfnamefont {R.~L.}\ \bibnamefont {Walsworth}}, \bibinfo {author} {\bibfnamefont {H.}~\bibnamefont {Park}}, \ and\ \bibinfo {author} {\bibfnamefont {M.~D.}\ \bibnamefont {Lukin}},\ }\href {\doibase 10.1103/PhysRevLett.113.197601} {\bibfield  {journal} {\bibinfo  {journal} {Phys. Rev. Lett.}\ }\textbf {\bibinfo {volume} {113}},\ \bibinfo {pages} {197601} (\bibinfo {year} {2014})}\BibitemShut {NoStop}%
\bibitem [{\citenamefont {Xiao}\ and\ \citenamefont {Zhao}(2016)}]{NJP_XZ_1}%
  \BibitemOpen
  \bibfield  {author} {\bibinfo {author} {\bibfnamefont {X.}~\bibnamefont {Xiao}}\ and\ \bibinfo {author} {\bibfnamefont {N.}~\bibnamefont {Zhao}},\ }\href {\doibase 10.1088/1367-2630/18/10/103022} {\bibfield  {journal} {\bibinfo  {journal} {New J. Phys.}\ }\textbf {\bibinfo {volume} {18}} (\bibinfo {year} {2016}),\ 10.1088/1367-2630/18/10/103022}\BibitemShut {NoStop}%
\bibitem [{\citenamefont {Dakir}\ \emph {et~al.}(2023)\citenamefont {Dakir}, \citenamefont {Slaoui}, \citenamefont {Mohamed}, \citenamefont {Laamara},\ and\ \citenamefont {Eleuch}}]{SR_DSM_1}%
  \BibitemOpen
  \bibfield  {author} {\bibinfo {author} {\bibfnamefont {Y.}~\bibnamefont {Dakir}}, \bibinfo {author} {\bibfnamefont {A.}~\bibnamefont {Slaoui}}, \bibinfo {author} {\bibfnamefont {A.-B.~A.}\ \bibnamefont {Mohamed}}, \bibinfo {author} {\bibfnamefont {R.~A.}\ \bibnamefont {Laamara}}, \ and\ \bibinfo {author} {\bibfnamefont {H.}~\bibnamefont {Eleuch}},\ }\href {\doibase 10.1038/s41598-023-46396-2} {\bibfield  {journal} {\bibinfo  {journal} {Sci. Rep.}\ }\textbf {\bibinfo {volume} {13}},\ \bibinfo {pages} {20526} (\bibinfo {year} {2023})}\BibitemShut {NoStop}%
\bibitem [{Qnami()}]{Qnami}%
  \BibitemOpen
  Qnami,\ \href@noop {} {\enquote {\bibinfo {title} {Quantum sensing leaders in nanoscale precision|qnami},}\ }\bibinfo {howpublished} {\url{https://qnami.ch/}} (\bibinfo {year} {2024})\BibitemShut {NoStop}%
\bibitem [{\citenamefont {Childress}\ \emph {et~al.}(2006{\natexlab{b}})\citenamefont {Childress}, \citenamefont {Dutt}, \citenamefont {Taylor}, \citenamefont {Zibrov}, \citenamefont {Jelezko}, \citenamefont {Wrachtrup}, \citenamefont {Hemmer},\ and\ \citenamefont {Lukin}}]{Childress_2006}%
  \BibitemOpen
  \bibfield  {author} {\bibinfo {author} {\bibfnamefont {L.}~\bibnamefont {Childress}}, \bibinfo {author} {\bibfnamefont {M.~V.~G.}\ \bibnamefont {Dutt}}, \bibinfo {author} {\bibfnamefont {J.~M.}\ \bibnamefont {Taylor}}, \bibinfo {author} {\bibfnamefont {A.~S.}\ \bibnamefont {Zibrov}}, \bibinfo {author} {\bibfnamefont {F.}~\bibnamefont {Jelezko}}, \bibinfo {author} {\bibfnamefont {J.}~\bibnamefont {Wrachtrup}}, \bibinfo {author} {\bibfnamefont {P.~R.}\ \bibnamefont {Hemmer}}, \ and\ \bibinfo {author} {\bibfnamefont {M.~D.}\ \bibnamefont {Lukin}},\ }\href {\doibase 10.1126/science.1131871} {\bibfield  {journal} {\bibinfo  {journal} {Science}\ }\textbf {\bibinfo {volume} {314}},\ \bibinfo {pages} {281} (\bibinfo {year} {2006}{\natexlab{b}})},\ \Eprint {http://arxiv.org/abs/https://www.science.org/doi/pdf/10.1126/science.1131871} {https://www.science.org/doi/pdf/10.1126/science.1131871} \BibitemShut {NoStop}%
\bibitem [{\citenamefont {Stepanov}\ and\ \citenamefont {Takahashi}(2016)}]{PRB_Stepanov_1}%
  \BibitemOpen
  \bibfield  {author} {\bibinfo {author} {\bibfnamefont {V.}~\bibnamefont {Stepanov}}\ and\ \bibinfo {author} {\bibfnamefont {S.}~\bibnamefont {Takahashi}},\ }\href {\doibase 10.1103/PhysRevB.94.024421} {\bibfield  {journal} {\bibinfo  {journal} {Phys. Rev. B}\ }\textbf {\bibinfo {volume} {94}},\ \bibinfo {pages} {024421} (\bibinfo {year} {2016})}\BibitemShut {NoStop}%
\bibitem [{\citenamefont {Loubser}\ and\ \citenamefont {van Wyk}(1978)}]{RPP_Loubser_1}%
  \BibitemOpen
  \bibfield  {author} {\bibinfo {author} {\bibfnamefont {J.~H.~N.}\ \bibnamefont {Loubser}}\ and\ \bibinfo {author} {\bibfnamefont {J.~A.}\ \bibnamefont {van Wyk}},\ }\href {\doibase 10.1088/0034-4885/41/8/002} {\bibfield  {journal} {\bibinfo  {journal} {Rep. Prog. Phys.}\ }\textbf {\bibinfo {volume} {41}},\ \bibinfo {pages} {1201} (\bibinfo {year} {1978})}\BibitemShut {NoStop}%
\bibitem [{\citenamefont {Baker}(1999)}]{Baker}%
  \BibitemOpen
  \bibfield  {author} {\bibinfo {author} {\bibfnamefont {J.~M.}\ \bibnamefont {Baker}},\ }\href {\doibase 10.1080/10420159908226267} {\bibfield  {journal} {\bibinfo  {journal} {Radiat. Eff. Defects Solids}\ }\textbf {\bibinfo {volume} {150}},\ \bibinfo {pages} {415} (\bibinfo {year} {1999})}\BibitemShut {NoStop}%
\bibitem [{\citenamefont {Peng}\ \emph {et~al.}(2019)\citenamefont {Peng}, \citenamefont {Biktagirov}, \citenamefont {Cho}, \citenamefont {Gerstmann},\ and\ \citenamefont {Takahashi}}]{JCP_Peng_1}%
  \BibitemOpen
  \bibfield  {author} {\bibinfo {author} {\bibfnamefont {Z.}~\bibnamefont {Peng}}, \bibinfo {author} {\bibfnamefont {T.}~\bibnamefont {Biktagirov}}, \bibinfo {author} {\bibfnamefont {F.~H.}\ \bibnamefont {Cho}}, \bibinfo {author} {\bibfnamefont {U.}~\bibnamefont {Gerstmann}}, \ and\ \bibinfo {author} {\bibfnamefont {S.}~\bibnamefont {Takahashi}},\ }\href {\doibase 10.1063/1.5085351} {\bibfield  {journal} {\bibinfo  {journal} {J. Chem. Phys.}\ }\textbf {\bibinfo {volume} {150}},\ \bibinfo {pages} {134702} (\bibinfo {year} {2019})}\BibitemShut {NoStop}%
\bibitem [{\citenamefont {Peng}\ \emph {et~al.}(2020)\citenamefont {Peng}, \citenamefont {Dallas},\ and\ \citenamefont {Takahashi}}]{JAP_Peng_1}%
  \BibitemOpen
  \bibfield  {author} {\bibinfo {author} {\bibfnamefont {Z.}~\bibnamefont {Peng}}, \bibinfo {author} {\bibfnamefont {J.}~\bibnamefont {Dallas}}, \ and\ \bibinfo {author} {\bibfnamefont {S.}~\bibnamefont {Takahashi}},\ }\href {\doibase 10.1063/5.0007599} {\bibfield  {journal} {\bibinfo  {journal} {J. Appl. Phys.}\ }\textbf {\bibinfo {volume} {128}},\ \bibinfo {pages} {054301} (\bibinfo {year} {2020})}\BibitemShut {NoStop}%
\bibitem [{\citenamefont {Norman}(2016)}]{RMP_Norman}%
  \BibitemOpen
  \bibfield  {author} {\bibinfo {author} {\bibfnamefont {M.~R.}\ \bibnamefont {Norman}},\ }\href {\doibase 10.1103/RevModPhys.88.041002} {\bibfield  {journal} {\bibinfo  {journal} {Rev. Mod. Phys.}\ }\textbf {\bibinfo {volume} {88}},\ \bibinfo {pages} {041002} (\bibinfo {year} {2016})}\BibitemShut {NoStop}%
\bibitem [{\citenamefont {Takagi}\ \emph {et~al.}(2019)\citenamefont {Takagi}, \citenamefont {Takayama}, \citenamefont {Jackeli}, \citenamefont {Khaliullin},\ and\ \citenamefont {Nagler}}]{Nature_Takagi_1}%
  \BibitemOpen
  \bibfield  {author} {\bibinfo {author} {\bibfnamefont {H.}~\bibnamefont {Takagi}}, \bibinfo {author} {\bibfnamefont {T.}~\bibnamefont {Takayama}}, \bibinfo {author} {\bibfnamefont {G.}~\bibnamefont {Jackeli}}, \bibinfo {author} {\bibfnamefont {G.}~\bibnamefont {Khaliullin}}, \ and\ \bibinfo {author} {\bibfnamefont {S.~E.}\ \bibnamefont {Nagler}},\ }\href {\doibase 10.1038/s42254-019-0038-2} {\bibfield  {journal} {\bibinfo  {journal} {Nat. Rev. Phys.}\ }\textbf {\bibinfo {volume} {1}},\ \bibinfo {pages} {264} (\bibinfo {year} {2019})}\BibitemShut {NoStop}%
\bibitem [{\citenamefont {Randall}\ \emph {et~al.}(2021)\citenamefont {Randall}, \citenamefont {Bradley}, \citenamefont {van~der Gronden}, \citenamefont {Galicia}, \citenamefont {Abobeih}, \citenamefont {Markham}, \citenamefont {Twitchen}, \citenamefont {Machado}, \citenamefont {Yao},\ and\ \citenamefont {Taminiau}}]{Science_Randall_1}%
  \BibitemOpen
  \bibfield  {author} {\bibinfo {author} {\bibfnamefont {J.}~\bibnamefont {Randall}}, \bibinfo {author} {\bibfnamefont {C.~E.}\ \bibnamefont {Bradley}}, \bibinfo {author} {\bibfnamefont {F.~V.}\ \bibnamefont {van~der Gronden}}, \bibinfo {author} {\bibfnamefont {A.}~\bibnamefont {Galicia}}, \bibinfo {author} {\bibfnamefont {M.~H.}\ \bibnamefont {Abobeih}}, \bibinfo {author} {\bibfnamefont {M.}~\bibnamefont {Markham}}, \bibinfo {author} {\bibfnamefont {D.~J.}\ \bibnamefont {Twitchen}}, \bibinfo {author} {\bibfnamefont {F.}~\bibnamefont {Machado}}, \bibinfo {author} {\bibfnamefont {N.~Y.}\ \bibnamefont {Yao}}, \ and\ \bibinfo {author} {\bibfnamefont {T.~H.}\ \bibnamefont {Taminiau}},\ }\href {\doibase 10.1126/science.abk0603} {\bibfield  {journal} {\bibinfo  {journal} {Science}\ }\textbf {\bibinfo {volume} {374}},\ \bibinfo {pages} {1474} (\bibinfo {year} {2021})}\BibitemShut {NoStop}%
\bibitem [{\citenamefont {Jones}\ \emph {et~al.}(2009)\citenamefont {Jones}, \citenamefont {Karlen}, \citenamefont {Fitzsimons}, \citenamefont {Ardavan}, \citenamefont {Benjamin}, \citenamefont {Briggs},\ and\ \citenamefont {Morton}}]{Science_Jonathan}%
  \BibitemOpen
  \bibfield  {author} {\bibinfo {author} {\bibfnamefont {J.~A.}\ \bibnamefont {Jones}}, \bibinfo {author} {\bibfnamefont {S.~D.}\ \bibnamefont {Karlen}}, \bibinfo {author} {\bibfnamefont {J.}~\bibnamefont {Fitzsimons}}, \bibinfo {author} {\bibfnamefont {A.}~\bibnamefont {Ardavan}}, \bibinfo {author} {\bibfnamefont {S.~C.}\ \bibnamefont {Benjamin}}, \bibinfo {author} {\bibfnamefont {G.~A.~D.}\ \bibnamefont {Briggs}}, \ and\ \bibinfo {author} {\bibfnamefont {J.~J.~L.}\ \bibnamefont {Morton}},\ }\href {\doibase 10.1126/science.1170730} {\bibfield  {journal} {\bibinfo  {journal} {Science}\ }\textbf {\bibinfo {volume} {324}},\ \bibinfo {pages} {1166} (\bibinfo {year} {2009})}\BibitemShut {NoStop}%
\end{thebibliography}%


\begin{thebibliography}{6}%
\makeatletter
\providecommand \@ifxundefined [1]{%
 \@ifx{#1\undefined}
}%
\providecommand \@ifnum [1]{%
 \ifnum #1\expandafter \@firstoftwo
 \else \expandafter \@secondoftwo
 \fi
}%
\providecommand \@ifx [1]{%
 \ifx #1\expandafter \@firstoftwo
 \else \expandafter \@secondoftwo
 \fi
}%
\providecommand \natexlab [1]{#1}%
\providecommand \enquote  [1]{``#1''}%
\providecommand \bibnamefont  [1]{#1}%
\providecommand \bibfnamefont [1]{#1}%
\providecommand \citenamefont [1]{#1}%
\providecommand \href@noop [0]{\@secondoftwo}%
\providecommand \href [0]{\begingroup \@sanitize@url \@href}%
\providecommand \@href[1]{\@@startlink{#1}\@@href}%
\providecommand \@@href[1]{\endgroup#1\@@endlink}%
\providecommand \@sanitize@url [0]{\catcode `\\12\catcode `\$12\catcode `\&12\catcode `\#12\catcode `\^12\catcode `\_12\catcode `\%12\relax}%
\providecommand \@@startlink[1]{}%
\providecommand \@@endlink[0]{}%
\providecommand \url  [0]{\begingroup\@sanitize@url \@url }%
\providecommand \@url [1]{\endgroup\@href {#1}{\urlprefix }}%
\providecommand \urlprefix  [0]{URL }%
\providecommand \Eprint [0]{\href }%
\providecommand \doibase [0]{http://dx.doi.org/}%
\providecommand \selectlanguage [0]{\@gobble}%
\providecommand \bibinfo  [0]{\@secondoftwo}%
\providecommand \bibfield  [0]{\@secondoftwo}%
\providecommand \translation [1]{[#1]}%
\providecommand \BibitemOpen [0]{}%
\providecommand \bibitemStop [0]{}%
\providecommand \bibitemNoStop [0]{.\EOS\space}%
\providecommand \EOS [0]{\spacefactor3000\relax}%
\providecommand \BibitemShut  [1]{\csname bibitem#1\endcsname}%
\let\auto@bib@innerbib\@empty
\bibitem [{\citenamefont {Schweiger}\ and\ \citenamefont {Jeschke}(2001)}]{principles_pepr}%
  \BibitemOpen
  \bibfield  {author} {\bibinfo {author} {\bibfnamefont {A.~A.}\ \bibnamefont {Schweiger}}\ and\ \bibinfo {author} {\bibfnamefont {G.}~\bibnamefont {Jeschke}},\ }\href@noop {} {\emph {\bibinfo {title} {Principles of pulse electron paramagnetic resonance}}}\ (\bibinfo  {publisher} {Oxford University Press},\ \bibinfo {address} {Oxford, UK ;},\ \bibinfo {year} {2001})\ \bibinfo {note} {includes bibliographical references (p. [542]-571) and index.}\BibitemShut {Stop}%
\bibitem [{\citenamefont {Mitrikas}\ and\ \citenamefont {Prokopiou}(2015)}]{Mitrikas_2015}%
  \BibitemOpen
  \bibfield  {author} {\bibinfo {author} {\bibfnamefont {G.}~\bibnamefont {Mitrikas}}\ and\ \bibinfo {author} {\bibfnamefont {G.}~\bibnamefont {Prokopiou}},\ }\href {\doibase 10.1016/j.jmr.2015.03.002} {\bibfield  {journal} {\bibinfo  {journal} {JOURNAL OF MAGNETIC RESONANCE}\ }\textbf {\bibinfo {volume} {254}},\ \bibinfo {pages} {75} (\bibinfo {year} {2015})}\BibitemShut {NoStop}%
\bibitem [{\citenamefont {Pham}\ \emph {et~al.}(2016)\citenamefont {Pham}, \citenamefont {DeVience}, \citenamefont {Casola}, \citenamefont {Lovchinsky}, \citenamefont {Sushkov}, \citenamefont {Bersin}, \citenamefont {Lee}, \citenamefont {Urbach}, \citenamefont {Cappellaro}, \citenamefont {Park}, \citenamefont {Yacoby}, \citenamefont {Lukin},\ and\ \citenamefont {Walsworth}}]{Pham_2016}%
  \BibitemOpen
  \bibfield  {author} {\bibinfo {author} {\bibfnamefont {L.~M.}\ \bibnamefont {Pham}}, \bibinfo {author} {\bibfnamefont {S.~J.}\ \bibnamefont {DeVience}}, \bibinfo {author} {\bibfnamefont {F.}~\bibnamefont {Casola}}, \bibinfo {author} {\bibfnamefont {I.}~\bibnamefont {Lovchinsky}}, \bibinfo {author} {\bibfnamefont {A.~O.}\ \bibnamefont {Sushkov}}, \bibinfo {author} {\bibfnamefont {E.}~\bibnamefont {Bersin}}, \bibinfo {author} {\bibfnamefont {J.}~\bibnamefont {Lee}}, \bibinfo {author} {\bibfnamefont {E.}~\bibnamefont {Urbach}}, \bibinfo {author} {\bibfnamefont {P.}~\bibnamefont {Cappellaro}}, \bibinfo {author} {\bibfnamefont {H.}~\bibnamefont {Park}}, \bibinfo {author} {\bibfnamefont {A.}~\bibnamefont {Yacoby}}, \bibinfo {author} {\bibfnamefont {M.}~\bibnamefont {Lukin}}, \ and\ \bibinfo {author} {\bibfnamefont {R.~L.}\ \bibnamefont {Walsworth}},\ }\href {\doibase 10.1103/PhysRevB.93.045425} {\bibfield  {journal} {\bibinfo  {journal} {Phys. Rev. B}\ }\textbf {\bibinfo {volume} {93}},\ \bibinfo {pages} {045425}
  (\bibinfo {year} {2016})}\BibitemShut {NoStop}%
\bibitem [{\citenamefont {Laraoui}\ \emph {et~al.}(2013)\citenamefont {Laraoui}, \citenamefont {Dolde}, \citenamefont {Burk}, \citenamefont {Reinhard}, \citenamefont {Wrachtrup},\ and\ \citenamefont {Meriles}}]{Laraoui_2013}%
  \BibitemOpen
  \bibfield  {author} {\bibinfo {author} {\bibfnamefont {A.}~\bibnamefont {Laraoui}}, \bibinfo {author} {\bibfnamefont {F.}~\bibnamefont {Dolde}}, \bibinfo {author} {\bibfnamefont {C.}~\bibnamefont {Burk}}, \bibinfo {author} {\bibfnamefont {F.}~\bibnamefont {Reinhard}}, \bibinfo {author} {\bibfnamefont {J.}~\bibnamefont {Wrachtrup}}, \ and\ \bibinfo {author} {\bibfnamefont {C.~A.}\ \bibnamefont {Meriles}},\ }\href {\doibase 10.1038/ncomms2685} {\bibfield  {journal} {\bibinfo  {journal} {Nature Communications}\ }\textbf {\bibinfo {volume} {4}},\ \bibinfo {pages} {1651} (\bibinfo {year} {2013})}\BibitemShut {NoStop}%
\bibitem [{\citenamefont {Felton}\ \emph {et~al.}(2009)\citenamefont {Felton}, \citenamefont {Edmonds}, \citenamefont {Newton}, \citenamefont {Martineau}, \citenamefont {Fisher}, \citenamefont {Twitchen},\ and\ \citenamefont {Baker}}]{Felton_2009}%
  \BibitemOpen
  \bibfield  {author} {\bibinfo {author} {\bibfnamefont {S.}~\bibnamefont {Felton}}, \bibinfo {author} {\bibfnamefont {A.~M.}\ \bibnamefont {Edmonds}}, \bibinfo {author} {\bibfnamefont {M.~E.}\ \bibnamefont {Newton}}, \bibinfo {author} {\bibfnamefont {P.~M.}\ \bibnamefont {Martineau}}, \bibinfo {author} {\bibfnamefont {D.}~\bibnamefont {Fisher}}, \bibinfo {author} {\bibfnamefont {D.~J.}\ \bibnamefont {Twitchen}}, \ and\ \bibinfo {author} {\bibfnamefont {J.~M.}\ \bibnamefont {Baker}},\ }\href {\doibase 10.1103/PhysRevB.79.075203} {\bibfield  {journal} {\bibinfo  {journal} {Phys. Rev. B}\ }\textbf {\bibinfo {volume} {79}},\ \bibinfo {pages} {075203} (\bibinfo {year} {2009})}\BibitemShut {NoStop}%
\bibitem [{\citenamefont {Smeltzer}\ \emph {et~al.}(2011)\citenamefont {Smeltzer}, \citenamefont {Childress},\ and\ \citenamefont {Gali}}]{Smeltzer_2011}%
  \BibitemOpen
  \bibfield  {author} {\bibinfo {author} {\bibfnamefont {B.}~\bibnamefont {Smeltzer}}, \bibinfo {author} {\bibfnamefont {L.}~\bibnamefont {Childress}}, \ and\ \bibinfo {author} {\bibfnamefont {A.}~\bibnamefont {Gali}},\ }\href {\doibase 10.1088/1367-2630/13/2/025021} {\bibfield  {journal} {\bibinfo  {journal} {New Journal of Physics}\ }\textbf {\bibinfo {volume} {13}},\ \bibinfo {pages} {025021} (\bibinfo {year} {2011})}\BibitemShut {NoStop}%
\end{thebibliography}%

\end{document}


\title{Supplementary Information for Detection of Electron Paramagnetic Resonance of Two Electron Spins Using a Single NV Center in Diamond}

\author{Yuhang Ren}
\email{yuhangre@usc.edu}
\affiliation{Department of Physics and Astronomy, University of Southern California, Los Angeles, 90089. CA, USA}

\author{Susumu Takahashi}
\email{susumuta@usc.edu}
\affiliation{Department of Physics and Astronomy, University of Southern California, Los Angeles, 90089. CA, USA}
\affiliation{Department of Chemistry, University of Southern California, Los Angeles, 90089. CA, USA}

\maketitle

\renewcommand{\figurename}{Fig.}
\renewcommand{\thefigure}{S\arabic{figure}}
\renewcommand{\thesection}{S\arabic{section}}
\renewcommand{\thesubsection}{S\arabic{section}.\arabic{subsection}}
\renewcommand{\theequation}{S\arabic{equation}}
\renewcommand{\thetable}{S\arabic{table}}

\section{Electron spin echo envelope modulation}

Electron spin echo envelope modulation (ESEEM) is an effect often observed in EPR experiments. When the hyperfine coupling between electron spin to nearby nuclear spins causes state mixing, oscillation in the envelope of spin echo arises due to the interference between perturbed EPR transitions. ESEEM spectroscopy is a powerful tool for proving a hyperfine coupling between an electron and surrounding nuclear spins. In the present study, the ESEEM signals on the NV center can be caused by bath $^{13}$C nuclear spins and individual nuclear spins, such as the $^{14}$N spin in the NV center and a neighboring $^{13}$C nuclear spin. Here, we discuss the ESEEM signals from those contributions.

We begin our discussion by considering an ESEEM signal due to coupling to a neighboring individual nuclear spin. We consider a system with an electron and nuclear spin and the hyperfine coupling of $A_{\parallel} Sz Iz + A_{\bot} (Sx Ix + Sy Iy)$. By considering that the $g$-tensor of the electron spin is isotropic and by applying the secular approximation, the system Hamiltonian is given by \cite{principles_pepr},
\begin{equation}
\label{eqs1}
    H_0 = \omega_S S_z + \omega_I I_z + AS_z I_z + B S_z I_x,
\end{equation}
where the first and second terms are the Zeeman terms of the electron and nuclear spins, and $\omega_S = \gamma_e B_0$ and $\omega_I = \gamma_n B_0$. $B_0$ is the static magnetic field. $\gamma_e$ is the gyromagnetic ratio of the NV center (28 GHz/T) and $\gamma_n$ is the gyromagnetic ratio of the nuclear spin (10.708 MHz/T for $^{13}$C and 3.077 MHz/T for $^{14}$N). $A$ and $B$ describe the secular and pseudo-secular hyperfine couplings. They are related to $A_{\parallel}$ and $A_{\bot}$ with the following expression:
\begin{equation}
    A = A_{\parallel} cos^2\theta + A_{\bot} sin^2\theta
\end{equation}
and
\begin{equation}
    B = (A_{\parallel} - A_{\bot})sin\theta cos\theta,
\end{equation}
where $\theta$ is the angle between the electron-nuclear spin axis and the static magnetic field. As described earlier, the polarization from the electron to the nuclear spin occurs under the Hamiltonian in Eqn.~\ref{eqs1}, and the coherence state is modulated by the following \cite{Mitrikas_2015},
\begin{multline}
\label{eqs4}
    V(\tau) = 1 + \frac{sin^2(N\varphi/4)}{sin^2(\varphi/2)} \left[-\frac{3k}{4} + \frac{k}{2}(cos\omega_{\alpha}\tau + cos\omega_{\beta}\tau) + \frac{k}{4}(cos2\omega_{\alpha}\tau + cos2\omega_{\beta}\tau) \right. \\ + \frac{k(\mu-\lambda)}{2}(cos\omega_+\tau - cos\omega_-\tau) + \frac{k\mu}{4}cos2\omega_+\tau + \frac{k\lambda}{4}cos2\omega_-\tau \\ \left. -\frac{k\mu}{2}[cos(\omega_+ + \omega_{\alpha})\tau + cos(\omega_+ + \omega_{\beta})\tau] -\frac{k\lambda}{2}[cos(\omega_- + \omega_{\alpha})\tau + cos(\omega_- - \omega_{\beta})\tau] \right]
\end{multline}
where N is the number of $\pi$ pulses in the dynamical decoupling sequence, and in our CPMG-8 sequence $N=8$. The modulation is characterized by the basic frequencies, $\omega_{\alpha}$ and $\omega_{\beta}$,
\begin{equation}
\label{eq5}
    \omega_{\alpha,\beta} = \sqrt{(\omega_I + m_S^{\alpha,\beta}A)^2+(m_S^{\alpha,\beta}B)^2}
\end{equation}
and the coefficients $\mu$ and $\lambda$, 
\begin{equation}
    \mu = cos^2\left[ [tan^{-1}(\frac{-m_S^{\alpha}B}{A + m_S^{\alpha}\omega_I}) - tan^{-1}(\frac{-m_S^{\beta}B}{A + m_S^{\beta}\omega_I})]/2 \right]
\end{equation}
and 
\begin{equation}
    \lambda = sin^2\left[ [tan^{-1}(\frac{-m_S^{\alpha}B}{A + m_S^{\alpha}\omega_I}) - tan^{-1}(\frac{-m_S^{\beta}B}{A + m_S^{\beta}\omega_I})]/2 \right].
\end{equation}
$m_S^{\alpha,\beta}$ values are spin quantum numbers for the electron spin. $\varphi = 2cos^{-1}(\mu cos\omega_+\tau + \lambda cos\omega_-\tau)$. The combination frequencies are
\begin{equation}
    \omega_{\pm} = \omega_{\alpha} \pm \omega_{\beta}.
\end{equation}
In addition, the modulation depth ($k$) is given as
\begin{equation}
\label{eq9}
    k = \left( \frac{B\omega_I(m_S^{\alpha}-m_S^{\beta})}{\omega_{\alpha}\omega_{\beta}} \right)^2.
\end{equation}
In the present case, the electron spin is the NV center with S = 1, and the transition between $\ket{m_S = 0}$ and $\ket{m_S = 1}$ is selectively excited, therefore, $m_S^{\alpha} = 0$ and $m_S^{\beta} = 1$. Then, Eqn.~\ref{eq5}-\ref{eq9} can be rewritten by,
\begin{gather}
    \notag \omega_{\alpha} = \omega_I, \omega_{\beta} = \sqrt{(\omega_I + A)^2+B^2}, \\
    \notag \mu = cos^2 \left[(tan^{-1}(\frac{-B}{A + \omega_I})/2 \right], \lambda = sin^2 \left[(tan^{-1}(\frac{-B}{A + \omega_I})/2 \right], \\
    \notag \omega_{\pm} = \omega_I \pm \sqrt{(\omega_I + A)^2+B^2}, \\
    k = \left( \frac{B}{\sqrt{(\omega_I + A)^2+B^2}} \right)^2. 
\end{gather}

\begin{figure}[ht]
\includegraphics[width=0.8\textwidth]{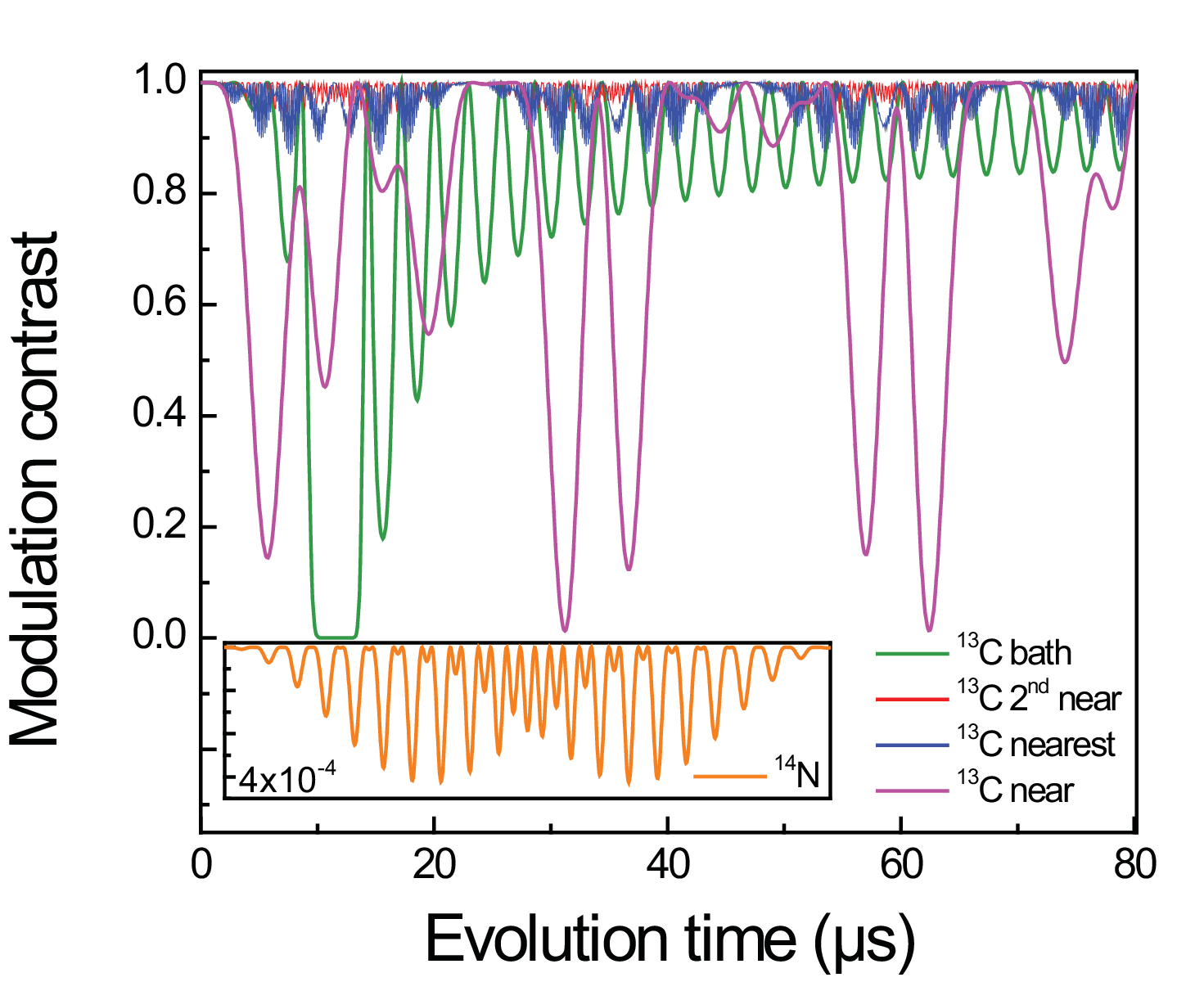}
\caption{\label{SI1}Simulation of the ESEEM based on the magnetic field and tilt angle found for NV1 and hyperfine interaction strength. Decoherence is not considered in this simulation. The modulation depth from $^{14}$N is negligible ($\sim$ 10$^{-4}$) compared to those from $^{13}$C near neighbors and bath, which is shown in the inlet. }
\end{figure}

Next, we consider the ESEEM signal from bath $^{13}$C spins. In this case, the modulation is caused by couplings to ensembles of nuclear spins, and the modulation depth depends on the strength of the magnetic field fluctuations caused by the bath nuclear spins. By considering the total magnetic fields from surrounding nuclear spins and an effect of the pulse sequence, the modulation on the coherence state is given by \cite{Pham_2016},
\begin{equation}
\label{eqs9}
    C(\tau) = exp\left[ -\frac{2}{\pi^2}\gamma_e^2 B^2_{RMS} K(N\tau)\right],
\end{equation}
where $\gamma_e \sim 2\pi \times$28 GHz/T is the NV's gyromagnetic ratio. $B_{RMS}$ is the RMS magnetic field produced by bath $^{13}$C spins (the natural abundance of 1.1\%) and is calculated to be 4 $\mu$T \cite{Laraoui_2013}. The functional $K(\tau) \sim (N\tau)^2 sinc^2[\frac{N\tau}{2}(\omega_I - \frac{\pi}{\tau})]$ describes a filtering effect and $N$ is number of $\pi$ pulses in the dynamical decoupling sequence.

\begin{table}[h]
\label{tables1}
\caption{A table listing the hyperfine coupling of the nuclear spin species calculated in the simulation. *$^{14}$N is an $I=1$ spin, but the formula for modulation depth is the same for $I=1/2$ spins.}
\begin{tabular}{ |p{4cm}||p{3cm}|p{3cm}|p{3cm}|  }
 \hline
Nuclear spin & $A$ (2$\pi\times$MHz) & $B$ (2$\pi\times$MHz) \\
 \hline
 $^{13}$C (nearest site) \cite{Felton_2009}& 921.080 & -229.240 \\
 $^{13}$C (2$^{nd}$ near site) \cite{Smeltzer_2011}& 94.288 & -15.498 \\
 $^{13}$C (near site)& 0.314 & 2.827 \\
 $^{14}$N* \cite{Felton_2009}& -13.459 & 0.214 \\
 \hline
\end{tabular}
\end{table}

Finally, we discuss the simulation of the NV ESEEM signals using Eqn.~\ref{eqs4} and~\ref{eqs9}. In the simulation, we consider the nuclear spins and hyperfine constants shown in Table~\ref{tables1}. As can be seen, we consider the 1$^{st}$ and 2$^{nd}$ nearest neighbor $^{13}$C spins, neighbor $^{13}$C spin with $A=0.314$ (2$\pi\times$MHz) and $B=2.827$ (2$\pi\times$MHz), $^{14}$N spin in the NV center and the bath $^{13}$C spins. The result is shown in Fig.~\ref{SI1}. The experimentally observed ESEEM effects in CPMG measurement of NV1 are explained by the simulated dominant effect from bath $^{13}$C spins and a neighbor $^{13}$C spin.

\section{NV centers and NV-EPR measurements}

In the present study, we investigated four single NV centers, including one presented in the main manuscript (NV1-4). Here, we summarized the experimental results of NV2-4. In our NV-EPR experiment, we start with the characterization of the magnetic field and the coherence manipulation capability of the NV centers. After identifying a single NV center, we perform a pulsed ODMR measurement. The criteria for screening out candidates for later measurements is that the NV center needs to have a large ODMR contrast, which will be proportional to the NV-EPR signal contrast. To maintain the experiment time to be reasonable (not longer than a few hours), we look for NV centers with an ODMR contrast larger than 10\% in characterization. 

\subsection{NV2 results}

\begin{figure}[ht]
\includegraphics[width=0.9\textwidth]{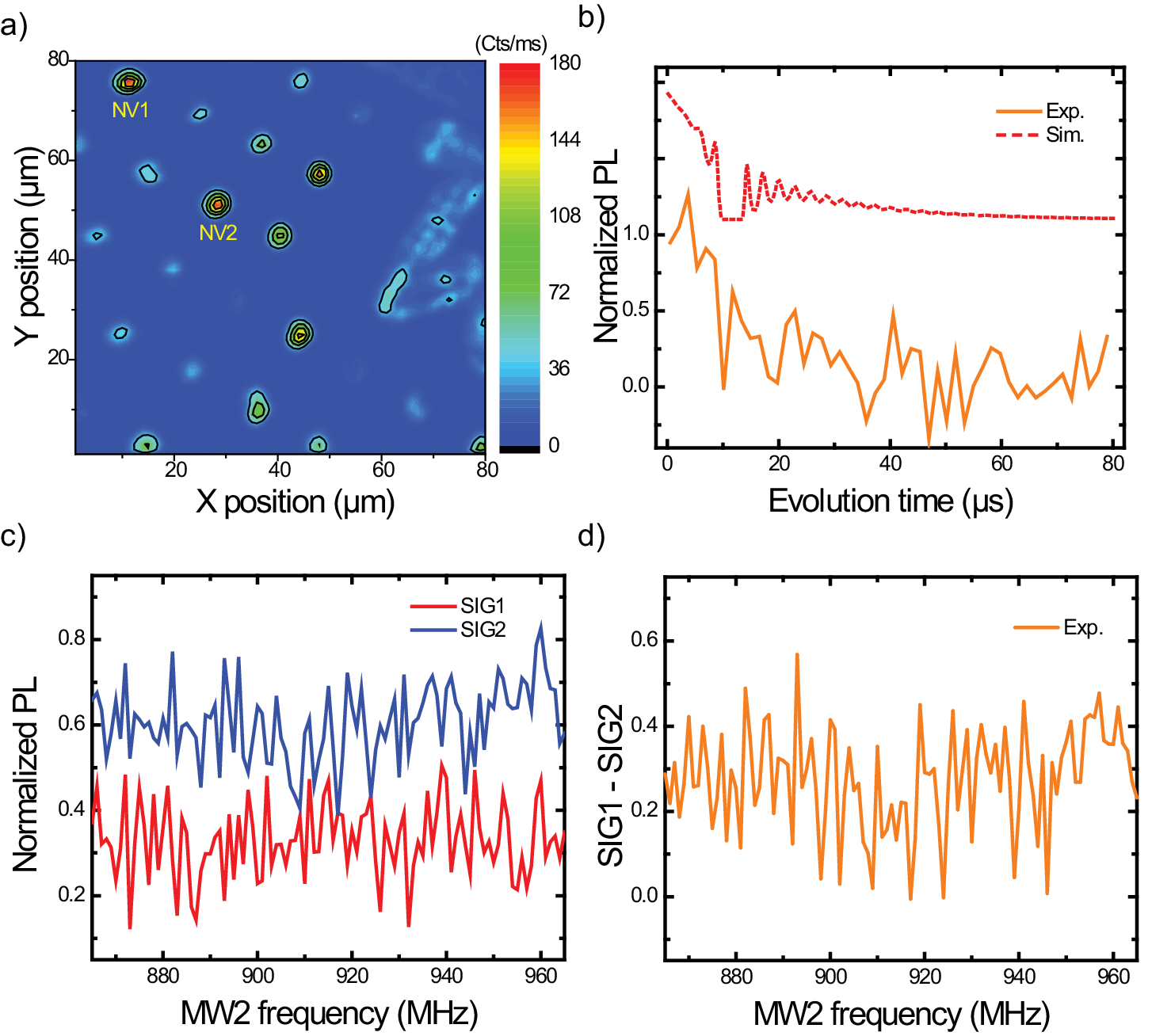}
\caption{\label{SI2}Characterization and NV-EPR measurement data of NV2. (a) PL image, showing the relative position of NV2 to NV1. (b) Normalized CPMG-8 data. The data is fitted to an exponential decay and extracted $T_2=17\pm2$ $\mu$s. MW1 $\pi/2$, $\pi$ and $3\pi/2$ pulse lengths are 48, 96 and 142 ns. Measurement time is around 28 mins. N = 195,000. (c) NV-EPR data. The same CPMG-DEER pulse sequence was used. $\tau$ time is set to 1.4 $\mu$s, and the total sensing time is 22.784 $\mu$s. The measurement time for this spectrum is around 330 minutes. N = 1,330,000. (d) The difference (SIG2-SIG1) of the data shows no signal with SNR larger than 1.}
\end{figure}

NV2 is found a few structure sites away from NV1, roughly 20 $\mu$m away (PL image shown in Fig.~\ref{SI2}(a)). The ODMR contrast is 16.6\%. The magnetic field is characterized to be 32.58 $\pm$ 0.02 mT from the same analysis from the main text. The spin coherence time ($T_2$) is measured with the CPMG-8 sequence and is extracted to be $17\pm2$ $\mu$s from a fit to an exponential decay simulation including the $^{13}$C bath spins effect (shown in Fig.~\ref{SI2}(b)). No clear sign of nearby $^{13}$C ESEEM effects was observed. In the NV-EPR measurement (data shown in Fig.~\ref{SI2}(c)), the resultant coherent state is mapped into the $\ket{m_s=0}$ state (SIG1) and the $\ket{m_s=1}$ state (SIG2). The pulse sequence for characterization and NV-EPR measurements are the
same as described in the main text. Using these two signal channels, we improved the signal-to-noise ratio (SNR) by removing systematic noise and doubling signal contrast. PL intensity is normalized using the same method described in the main text (shown in Fig.~\ref{SI2}(d)). With a 5.5-hour data averaging, we did not resolve a signal with an SNR larger than one. 

\begin{figure}[ht]
\includegraphics[width=0.8\textwidth]{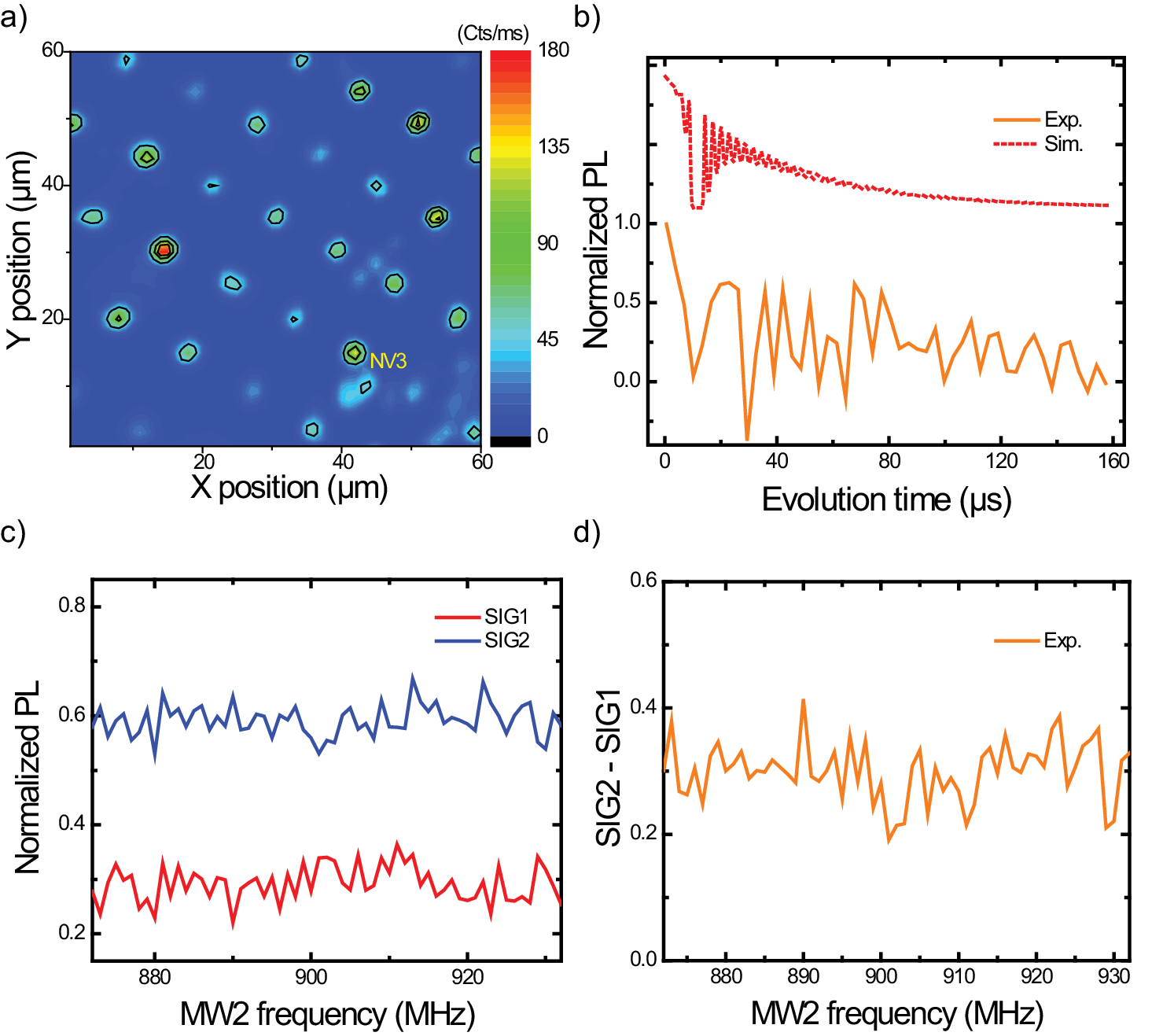}
\caption{\label{SI3}Characterization and NV-EPR measurement data of NV3. (a) PL image of NV3. (b) Normalized CPMG-8 data. We extracted $T_2=40\pm4$ $\mu$s. MW1 $\pi/2$, $\pi$ and $3\pi/2$ pulse lengths are 52, 104 and 156 ns. Measurement time is around 26 mins. N = 125,000. (c) NV-EPR data. $\tau$ time is set to 4.2 $\mu$s, and the total sensing time is 67.584 $\mu$s. The measurement time for this spectrum is around 590 minutes. N = 2,500,000. (d) The difference (SIG2-SIG1) of the CPMG-DEER data.}
\end{figure}

\subsection{NV3 results}

NV3 is found at a site far from NV1 and NV2 (PL image shown in Fig.~\ref{SI3}(a)). The ODMR contrast is 12.7\%. The magnetic field is characterized to be 32.27 $\pm$ 0.04 mT with a tilt angle $\theta = 5 \pm 1$ degrees. The spin coherence time ($T_2$) is extracted to be $40\pm4$ $\mu$s from a fit to an exponential decay simulation including the $^{13}$C bath spins effect (shown in Fig.~\ref{SI3}(b)). We did not resolve a signal with an SNR larger than one in the 9.8 hours of signal averaging (shown in Fig.~\ref{SI2}(c) and (d)).

\begin{figure}[ht]
\includegraphics[width=0.8\textwidth]{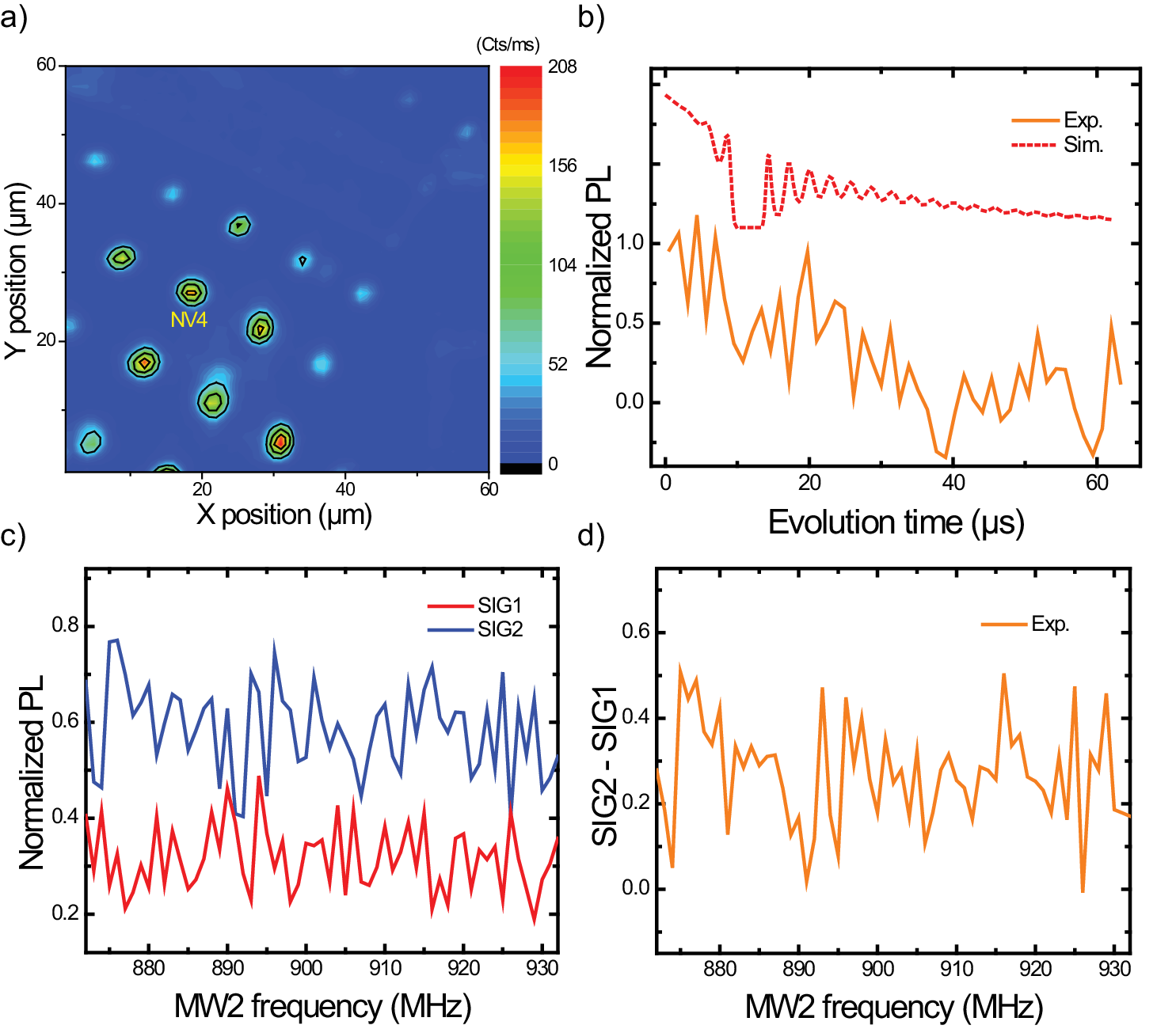}
\caption{\label{SI4}Characterization and NV-EPR measurement data of NV4. (a) PL image of NV4. (b) Normalized CPMG-8 data. We extracted $T_2=24\pm3$ $\mu$s. MW1 $\pi/2$, $\pi$ and $3\pi/2$ pulse lengths are 52, 104 and 156 ns. Measurement time is around 13 minutes. N = 100,000. (c) NV-EPR data. $\tau$ time is set to 1.6 $\mu$s, and the total sensing time is 25.984 $\mu$s. The measurement time for this spectrum is around 41 minutes. N = 265,000. (d) The difference (SIG2-SIG1) of the CPMG-DEER data.}
\end{figure}

\subsection{NV4 results}

NV4 is also found in a different site (PL image shown in Fig.~\ref{SI3}(a)). The ODMR contrast is 13.4\%. The magnetic field is characterized to be 32.24$\pm$ 0.04 mT with a tilt angle $\theta = 4 \pm 1$ degrees. The spin coherence time ($T_2$) is measured to be $24\pm3$ $\mu$s from a fit to an exponential decay simulation including the $^{13}$C bath spins effect(shown in Fig.~\ref{SI4}(b)). We did not resolve a signal with an SNR larger than one in 1 hour (shown in Fig.~\ref{SI4}(c) and (d)).

\bibliography{bibliographySI}